\documentclass[preprint,journal]{vgtc}            % preprint (journal style)

\onlineid{1261}

\vgtccategory{Research}

\title{Cohort-based Semantic Labeling: AI-Enabled Recovery of Visualization Semantics from Deployed SVGs}

\author{%
  Jeongah Lee,
  Hima Varshini Surisetty, 
  Durga Nirmaleswaran,
  Jahnavi Sharma,
  Srikiran Kavuri,\\
  Narges Mahyar, and 
  Ali Sarvghad
}

\authorfooter{
  \item
    Jeongah Lee is with the University of Massachusetts Amherst.
    E-mail: jeongahlee@umass.edu
  \item
    Hima Varshini Surisetty is with the University of Massachusetts Amherst.
    E-mail: hsurisetty@umass.edu
  \item
    Durga Nirmaleswaran is with the University of Massachusetts Amherst.
    E-mail: dnirmaleswar@umass.edu
  \item
    Jahnavi Sharma is with the University of Massachusetts Amherst.
    Email: jsharma@umass.edu 
  \item
    Srikiran Kavuri is with the University of Massachusetts Amherst.
    E-mail: skavuri@umass.edu
  \item
    Narges Mahyar is with Computer Science, City St George's, University of London.
    E-mail: narges.mahyar@city.ac.uk
  \item
    Ali Sarvghad is with Computer Science, City St George's, University of London.
    E-mail: ali.sarvghad@citystgeorges.ac.uk
}

 \abstract{
 Many web-based visualizations are deployed as Scalable Vector Graphics (SVG). While this format faithfully preserves visual appearance, it typically omits the higher-level semantic structure needed for machine interpretation. As a result, once visualizations are rendered and published, information about their components, roles, and encodings is no longer explicitly available, limiting downstream operations such as querying, accessibility augmentation, explanation, personalization, and post-deployment transformation. To address this gap, we introduce \CSL, an AI-enabled, multi-stage transformation pipeline for automatically recovering visualization semantics from deployed SVGs. \CSL\ enables post-deployment semantic recovery through two complementary mechanisms: (1) cohort-based decomposition and (2) hybrid semantic grounding. Cohort-based decomposition organizes heterogeneous SVG primitives into structurally coherent subsets that reduce the semantic assignment space, while hybrid semantic grounding combines model-based inference with deterministic structural validation and propagation to make labeling both context-sensitive and structurally anchored. \CSL\ produces Semantic SVG (SSVG), a semantically enriched representation in which SVG elements are annotated with graphical mark type, visualization role, and data role. We implemented \CSL\ as an end-to-end prototype and evaluated it on 102 SVG visualizations in terms of recovery accuracy, the contribution of cohorting, and labeling consistency. Across the full corpus, \CSL\ achieved global macro-averaged accuracies of 0.822 for mark type, 0.853 for visualization role, and 0.860 for data-role recovery. We also conducted an ablation study comparing the \CSL\ prototype with a non-cohort whole-chart baseline. The results showed that cohorting significantly improves semantic recovery and labeling accuracy (paired $t$-test: $t > 20$, $p < 0.001$; Cohen's $d > 2.0$). Finally, repeated labeling of a randomly selected SVG over 100 runs yielded mean agreement above 91.9\% across all three attributes, indicating highly consistent cohort-level semantic inference across repeated runs. Our results provide strong evidence that \CSL\ can transform deployed SVG visualizations from terminal graphical artifacts into machine-usable semantic representations, creating new opportunities for more accessible, adaptive, and user-steerable visualization systems.
 }


\usepackage{tabu}                      % only used for the table example
\usepackage{booktabs}                  % only used for the table example
\usepackage{lipsum}                    % used to generate placeholder text
\usepackage{mwe}                       % used to generate placeholder figures
\usepackage{ccicons}    
\usepackage{longtable}
\usepackage{array}
\usepackage{longtable}
\usepackage{makecell}
\usepackage{pdflscape}
\usepackage{tcolorbox}
\usepackage{tcolorbox}
\usepackage{listings}
\tcbuselibrary{breakable,skins,listings}
\usepackage{tabularx}
\usepackage{soul}
\usepackage{xcolor}
\usepackage{balance}
\usepackage{multirow}
\usepackage[table]{xcolor}
\usepackage{url}
\usepackage[pagebackref,bookmarks]{hyperref}

\definecolor{myred}{HTML}{A71D31}
\definecolor{myblue}{HTML}{284b63}

\definecolor{tagcolor}{HTML}{a71d31}
\definecolor{attrcolor}{HTML}{a71d31}
\definecolor{valcolor}{HTML}{284b63}

\renewcommand{\arraystretch}{1.2}
\newcommand{\CSL}{\texttt{\large CSL}}

\usepackage{mathptmx}  
\usepackage{amsmath} 
\begin{document}

%%%%%%%%%%%%%%%%%%%%%%%%%%%%%%%%%%%%%%%%%%%%%%%%%%%%%%%%%%%%%%%%
%%%%%%%%%%%%%%%%%%%%%% START OF THE PAPER %%%%%%%%%%%%%%%%%%%%%%
%%%%%%%%%%%%%%%%%%%%%%%%%%%%%%%%%%%%%%%%%%%%%%%%%%%%%%%%%%%%%%%%

%% The ``\maketitle'' command must be the first command after the
%% ``\begin{document}'' command. It prepares and prints the title block.
%% the only exception to this rule is the \firstsection command

\maketitle

\section{Introduction}

Modern web visualization frameworks such as D3~\cite{d3graphgallery, lin2025observable}, Vega~\cite{satyanarayan2016vega}, and Observable~\cite{d3observable} commonly produce graphics in Scalable Vector Graphics (SVG). During authoring, visualization specifications and data are compiled into rendered arrangements of graphical primitives such as rectangles, paths, lines, and text. Once deployed, the SVG becomes the artifact that users see and interact with in the browser~\cite{battle2018beagle}. Although visually expressive, rendered SVGs are often semantically sparse, they retain geometry, styling, and layout, but not explicit information about which elements correspond to data marks, axes, labels, legends, or other meaningful visualization components~\cite{li2022structure,snyder2025challenges,lee2025svg}.

Prior work has shown the value of structured semantic representations for visualization, including declarative grammars~\cite{narechania2020nl4dv}, scene abstractions, spatial constraint models for manipulating static visualizations~\cite{liu2024spatial}, curated design indices~\cite{hoque2020d3search}, and reverse-engineering pipelines for chart understanding~\cite{liu2023deplot}. These approaches demonstrate that semantically structured representations can enable interpretation, interaction, and transformation. 
However, these approaches do not fully address the post-deployment setting, where visualizations are encountered as rendered SVG artifacts rather than as authoring-time specifications. This missing semantic structure limits support for downstream operations such as programmatic querying, accessibility augmentation, explanation, and personalization~\cite{fan2023accessibility, sharif2021understanding}. For example, a person with accessibility needs viewing a bar chart in an online article may wish to enlarge axis labels, improve color contrast, identify marks above a threshold, or obtain a structured explanation of the chart~\cite{zong2022rich,thompson2023chart,jones2024customization,alam2023seechart}. Unless such capabilities are explicitly built in at authoring time, enabling them requires access to the semantic roles of visualization elements. If this organization could be recovered directly from the rendered artifact, post-deployment tools could support such requests by synthesizing interaction widgets or applying user-directed modifications directly to visualization components~\cite{vaithilingam2024dynavis}. Recovering such semantics is difficult. Deployed SVGs often contain varying primitives organized through nested groupings, inherited styling, and coordinate transformations, while authoring-time data bindings are no longer available~\cite{wang2025internsvg,chen2025svgenius}.

To address this problem, we introduce \texttt{Cohort-based Semantic Labeling (CSL)}, a multi-stage pipeline for recovering visualization semantics directly from deployed SVGs. 
At the heart of \CSL\  are (1) \emph{cohort-based decomposition}, and (2) \emph{hybrid semantic grounding}.
Cohorts are groups of SVG elements whose shared geometric, stylistic, and structural regularities make them plausible carriers of a common semantic interpretation. Cohorts serve as an intermediate unit of analysis between individual primitives and the full SVG. This is important because isolated elements often provide too little context for reliable interpretation, whereas whole-chart inference must reason over many heterogeneous structures simultaneously. By organizing the SVG into structurally coherent subsets, \CSL\  reduces the semantic assignment space, limits error propagation across unrelated elements, and makes semantic inference more stable and interpretable. The hybrid inference and grounding mechanism makes semantic labeling both context-sensitive and structurally anchored. The model performs inference over each cohort rather than over isolated elements or the full visualization at once. The inferred roles are then grounded in individual SVG elements through deterministic validation, and quantitative encodings are recovered via geometric measurement within reconstructed axis contexts.
Through this mechanism, \CSL\ transforms rendered SVG into \texttt{Semantic SVG (SSVG)}. SSVG is an enriched form of rendered SVG in which graphical primitives are augmented with explicit semantic roles and, where recoverable, structural relationships and value encodings. For example, a \textcolor{myred}{\lstinline!<rect .../>!} element may be identified not simply as a rectangle, but as a data mark serving as a bar in a bar chart.

To evaluate the feasibility and performance of \CSL, we implemented the pipeline in a prototype and tested it on a corpus of 102 SVG visualizations from diverse sources, covering 17,276 labeled elements. Across the corpus, \CSL\  achieved global macro-averaged accuracies of 0.822 for mark, 0.853 for role, and 0.860 for data-role recovery. 
We also conducted an ablation study comparing the \CSL\ prototype with a non-cohort whole-chart baseline, demonstrating that the cohort-based pipeline significantly outperforms a no-cohort variant across all three labeling dimensions (paired $t$-test: $t > 20$, $p < 0.001$; Cohen's $d > 2.0$). These results underscore the central role of cohorting in effective semantic recovery.
In addition, we conducted a 100-run labeling consistency assessment with a randomly selected SVG. This assessment yielded a mean agreement of over 91.9\% and a median of 92.0\% across mark, role, and data-role attributes.

Our results show that \CSL\  can recover element-level semantic labels with strong accuracy and stability.
An anonymized supplementary repository\footnote{\href{https://osf.io/nkxez/overview?view_only=3f12709420da4a2980607152d376381b}{Anonymous OSF supplementary materials}} provides the evaluation prototype, representative recovered SSVG outputs, and supporting evaluation materials for review.

The contributions of this paper are threefold:
\begin{enumerate}
    \item \texttt{Cohort-based Semantic Labeling (\CSL)}, a novel multi-stage AI-enabled pipeline for recovering semantic structure from rendered SVG.
    \item \texttt{Semantic SVG (SSVG)}, a recovered representation that makes element-level semantic roles explicit and can support post-deployment configurations.
    \item An end-to-end evaluation demonstrating the feasibility, tractability, and consistency of \CSL.
\end{enumerate}

In contrast to prior work on chart extraction, SVG deconstruction, or LLM-assisted adaptation, \CSL\ enables automated recovery of a persistent, general-purpose semantic representation directly from deployed SVG artifacts. By making element-level roles, structural relationships, and aspects of encoding explicit after deployment, \CSL\ lays the foundation for future systems that leverage recovered semantic structure to support post-deployment interaction, accessibility, querying, and other user-driven operations.

\section{Related Work} 
\label{sec:lr}

Prior work shows that semantic structure is central to visualization systems and that post hoc chart understanding is possible. Yet a gap remains between authoring-time semantic representations and the rendered artifacts encountered after deployment. This section positions our work in that gap and motivates recovering semantic structure directly from deployed SVG visualizations.

\subsection{Semantic structure in visualization systems} 

A central premise of our work is that visualization semantics are neither arbitrary nor incidental. Modern visualization systems are grounded in explicit theories and formalisms that define how data, marks, encodings, scales, and composition relate. The \emph{Grammar of Graphics} tradition, for example, models visualizations as structured compositions of independent components, enabling systematic reasoning about what a chart represents rather than how it appears \cite{wilkinson2011grammar, wilkinson2011ggplot2, wickham2010layered}. This perspective underlies both lower-level rendering toolkits such as D3, which expose the rendered SVG/ Document Object Model (DOM) substrate directly \cite{bostock2011d3}, and higher-level declarative grammars such as Vega-Lite, which specify mappings from data fields to visual channels together with composition and interaction semantics \cite{satyanarayan2016vega}.
% {kim2024erie} https://ieeexplore.ieee.org/abstract/document/10297592
% {kim2022cicero} https://www.tandfonline.com/doi/abs/10.1198/jcgs.2009.07098

Declarative ecosystems also make the semantic structure computationally actionable. Draco formalizes visualization design knowledge as constraints over structured chart representations and shows how such knowledge can support validation and optimization \cite{moritz2018formalizing}. 
Related work on declarative interaction and reactive dataflow likewise treats specifications as executable semantic programs rather than static drawings \cite{satyanarayan2016vega}. Tools such as VizLinter further demonstrate that structured representations enable automated checking, diagnosis, and repair of visualization designs \cite{chen2021vizlinter}, while Reactive Vega maintains semantic structure at runtime through a streaming dataflow architecture that keeps data bindings and encodings live throughout execution \cite{satyanarayan2015reactive}.

These systems clarify what semantic structure consists of in visualization, including marks, axes, legends, labels, layouts, and encodings, together with the relationships among them. However, these semantic programs are typically available only during authoring time. Once a visualization is exported and deployed as SVG, much of this explicit structure is no longer directly accessible \cite{satyanarayan2019critical}.

\subsection{Post hoc chart understanding and reverse engineering} 
\label{subsec:lr-reverse}

A substantial body of work shows that meaningful chart structure can be recovered after the fact. Early systems such as ReVision demonstrated automated classification, analysis, and redesign of rasterized chart images \cite{savva2011revision}, while Poco and Heer emphasized recovery of visual encodings as an intermediate representation for reverse engineering and retargeting charts \cite{poco2017reverse}. These image-based approaches establish the feasibility of post hoc chart understanding, but must contend with Optical Character Recognition (OCR), segmentation, and other perception challenges introduced by pixel-based inputs \cite{fu2020chartem, masry2022chartqa, methani2020plotqa}. Interactive systems such as ChartDetective further show the value of human-in-the-loop recovery for charts in the wild \cite{masson2023chartdetective}.

Within SVG-based settings, prior work has explored recovering structure from rendered visualizations, but under constrained assumptions. Harper and Agrawala extract data, marks, and data--attribute mappings from D3 visualization by accessing underlying data bindings in the Javascript runtime \cite{harper2014deconstructing, harper2017converting}, limiting applicability to specific frameworks and environments. 
More recent systems extend this direction toward reuse and interaction through task-specific or partial structure recovery \cite{zhu2019towards, hoque2020d3search, cui2021mixed}. Mystique deconstructs SVG charts into reusable components for authoring workflows \cite{chen2023mystique}, DIVI infers lightweight chart structure at runtime to support interaction \cite{snyder2023divi}, DataWink uses large multimodal models to recover reusable intermediate representations from SVG examples \cite{xie2025datawink}, and Ying et al. revive static SVGs into animated `Live Charts' with audio narrations \cite{ying2024reviving}.

These systems demonstrate that deployed SVGs retain usable structure, but they do not target fully automated recovery of a persistent, general-purpose semantic representation from arbitrary deployed SVG artifacts (Table ~\ref{tab:prior-comparison}). \CSL\ builds on this gap by recovering semantic roles, structural relationships, and quantitative encodings directly from deployed SVG, yielding a semantically enriched representation designed to support downstream operations beyond reuse, runtime interaction scaffolding, or example-based adaptation.

\subsection{Infrastructure for semantic recovery}
% Corpora, scene abstractions, and accessibility-oriented representations} 
\label{subsec:lr-corpora}

Recent work has created the infrastructure to make semantic recovery more measurable and reusable. VisAnatomy contributes a corpus of real-world SVG charts with multi-level semantic labels covering element roles, grouping structure, layout, and encodings \cite{chen2025visanatomy}. Such corpora are important for both training machine learning components and evaluating recovered semantic roles and relationships against reference annotations \cite{deng2022visimages}.
In parallel, scene-oriented representations provide useful abstractions for describing recovered structure. Manipulable Semantic Components (MSC), for example, propose an object model comprising semantic components such as marks, groups, layouts, and encodings, motivated by reverse engineering, authoring, and animation use cases \cite{liu2024manipulable}. More broadly, work surveying chart corpora emphasizes that corpus diversity and label granularity are key determinants of robustness for automated chart analysis systems \cite{chen2023corpora, wu2021ai4vis, dou2024hierarchically, battle2018beagle}.

Accessibility research provides an additional motivation for semantic recovery.  World Wide Web Consortium (W3C) specifications such as the Web Accessibility Initiative - Accessible Rich Internet Applications (WAI-ARIA) Graphics Module, Graphics-AAM, and SVG-AAM define how semantic chart structure should be exposed to assistive technologies through accessibility APIs \cite{w3c2018waiaria, w3c2018graphicsaam, w3c2026svgaam}. However, deployed SVG charts often lack the semantic markup needed to support such mappings. Systems such as AChart and AutoVizuA11y illustrate how semantically enriched chart representations can support accessible SVG output and screen-reader interaction \cite{andrews2024achart,duarte2024autovizua11y}. Similarly, work such as Tactile Vega-Lite and ChartFormer reinforces the importance of explicit chart structure as an output representation for accessibility-oriented downstream systems \cite{chen2025tactilevegalite,moured2024chartformer}.
These corpora, abstractions, and standards collectively clarify both what should be recovered and why recovering it matters. They supply the semantic targets, evaluation scaffolds, and downstream motivations that ground our work as a method for recovering semantic structure directly from deployed SVG artifacts.
%These corpora, abstractions, and standards help clarify both what should be recovered and why recovering it matters. They define target semantic categories and demonstrate the value of exposing chart structure beyond raw graphics.
%These efforts define the semantic targets, evaluation scaffolds, and downstream motivations that motivate our work as a method for recovering semantic structure directly from deployed SVG artifacts.

\begin{table*}[t]
\centering
\small
\renewcommand{\arraystretch}{1.0}
\begin{tabular}{p{5.4cm} p{12.0cm}}
\hline
\textbf{Prior Work} & \textbf{Recovered / Target Representation} \\
\hline
Beagle \cite{battle2018beagle}
& SVG chart instances with inferred chart-type labels for corpus-level analysis \\
ChartDetective \cite{masson2023chartdetective}
& Recovered underlying data from vector chart elements and their spatial structure \\
DataWink \cite{xie2025datawink}
& Semantic intermediate representation connecting primitive SVG elements to reusable templates \\
DIVI \cite{snyder2023divi}
& Standardized chart semantics inferred from SVG, including marks, axes, legends, and scales \\
Harper \& Agrawala \cite{harper2014deconstructing}
& Recovered data, marks, and data-to-visual-attribute mappings from D3/SVG charts \\
Hoque \& Agrawala \cite{hoque2020d3search}
& Searchable structural and style representation of D3 charts, including marks, encodings, axes, and style attributes \\
Mystique \cite{chen2023mystique}
& Layout-oriented semantic decomposition of SVG charts into groups, spatial relations, encodings, and constraints \\
Ying et al. \cite{ying2024reviving}
& Recovered chart data and visual encodings for animated and narrated Live Chart generation \\
\hline
\end{tabular}
\caption{Representative prior work on structured representations, semantic decomposition, and reverse engineering of visualizations that take SVG or vector-based chart artifacts as input. These works recover, index, or organize partial, task-specific chart structure, such as chart types, data, marks, encodings, metadata, layouts, or reusable components. In contrast, our approach targets automated recovery of a persistent, general-purpose semantic representation directly from deployed SVG artifacts, including element-level semantic roles, structural relationships, and aspects of quantitative encoding.}
\label{tab:prior-comparison}
\end{table*}

% Chartreuse \cite{cui2021mixed}
%& Reusable infographic chart model with data-binding and layout update strategies \\

% MSC \cite{liu2024manipulable}
%& Unified semantic scene representation of marks, groups, encodings, layouts, constraints, and operations \\

%VisAnatomy \cite{chen2025visanatomy}
%& Fine-grained semantic SVG corpus with element roles, group structure, layouts, and visual encodings \\

% VAID \cite{ying2024vaid}
%& Structured index of VA view designs covering analytical tasks and composite visual components \\

% Prior approaches variously target curated indexing, runtime extraction, mixed-initiative deconstruction, lightweight interaction metadata, or example-driven SVG adaptation.

\subsection{Gap and positioning} \label{subsec:lr-gap}

Previous literature supports three conclusions. First, semantic structure is foundational and well-defined in visualization research, as grammars and constraint systems make the marks, encodings, composition, and relationships that give charts meaning explicit \cite{wilkinson2011grammar,wilkinson2011ggplot2,satyanarayan2016vega,moritz2018formalizing,chen2021vizlinter}. 
Second, post hoc chart understanding is achievable, since prior work shows that meaningful structure can be inferred both from lossy raster inputs and from richer vector artifacts, although often with substantial ambiguity, restricted scope or reliance on interaction \cite{savva2011revision,poco2017reverse,chen2023mystique,masson2023chartdetective, jung2017chartsense}. 
% Third, there is now infrastructural support for measurable progress, with SVG corpora, scene abstractions, and accessibility standards providing target semantics and downstream motivations for semantic recovery 
Third, recent corpora, abstractions, and accessibility standards now provide the target
semantics and evaluation scaffolds needed to study recovery more systematically\cite{chen2025visanatomy,liu2024manipulable,w3c2018graphicsaam,w3c2026svgaam}.

% Despite these advances, an important gap remains between semantic representations available \emph{before} deployment and the rendered artifacts \emph{after} deployment. Existing work either assumes access to authoring-time specifications, depends on human-in-the-loop manual or interactive input to resolve ambiguity, focuses on raster settings where vector structure is absent, or operates on curated corpora rather than recovering semantics directly from deployed SVG artifacts \cite{ying2024vaid,chen2023mystique,savva2011revision, ying2024reviving, ye2023invvis}. 
% Our work addresses this gap by targeting automated recovery of semantic roles, structural relationships, and quantitative encodings directly from deployed SVG. 

Despite these advances, an important gap remains between pre-deployment semantic structure and the rendered artifacts available after deployment. Existing systems typically assume access to authoring-time specifications or runtime bindings, rely on manual or interactive disambiguation, operate on raster rather than deployed SVG inputs, or recover only partial, task-specific structure for extraction, reuse, restyling, or interaction scaffolding \cite{ying2024vaid,chen2023mystique,savva2011revision, ying2024reviving}. Our work addresses this gap by recovering a persistent, general-purpose semantic representation directly from deployed SVG artifacts. 
% Accordingly, \CSL\ recovers element-level semantic roles, structural relationships, and aspects of quantitative encoding to produce a machine-usable representation that supports downstream operations beyond any single task, including reuse, adaptation, and runtime interaction.

% Our work addresses this gap by targeting automated recovery of a persistent, general-purpose semantic representation directly from deployed SVG artifacts. 
% Accordingly, \CSL targets element-level semantic roles, structural relationships, and aspects of quantitative encoding, yielding a machine-usable representation intended to support downstream operations beyond any single task such as reuse, adaptation, or runtime interaction.

% \lstset{
%     language=XML,
%     basicstyle=\ttfamily\small,
%     breaklines=true,
%     breakatwhitespace=true,
%     tagstyle=\color{tagcolor},
%     keywordstyle=\color{attrcolor},
%     stringstyle=\color{valcolor},
%     showstringspaces=false,
%     columns=fullflexible
% }

\section{Cohort-based Semantic Labeling (CSL)}
\texttt{Cohort-based Semantic Labeling} is a multi-stage pipeline for recovering semantic structure from rendered SVG artifacts. Rather than treating SVG as a purely graphical output, \CSL\ reconstructs an enriched representation, \texttt{Semantic SVG (SSVG)}, in which graphical primitives are augmented with explicit semantic roles and structural relationships.

\CSL\ makes post-deployment semantic recovery tractable through two complementary foundations: cohort-based decomposition and hybrid semantic grounding. Because deployed SVG visualizations often contain hundreds or thousands of heterogeneous primitives, direct element-level semantic assignment is ambiguous and computationally unstable. \CSL\ addresses this by decomposing the visualization into structurally coherent \texttt{cohorts}, groups of elements that share geometric, stylistic, and structural fingerprints and likely participate in the same repeated visual pattern. These cohorts serve as intermediate \texttt{structural hypotheses} that reduce the semantic assignment space and provide stable units for interpretation.

Semantic interpretation then operates over cohorts rather than individual primitives. Multimodal models infer candidate semantic roles using both local cohort structure and global cues such as axis layout, repetition patterns, and chart organization, then ground these interpretations back to primitives through deterministic structural cues to produce a recovered \texttt{Semantic SVG} representation.

The \CSL\  pipeline proceeds through five stages.

\begin{enumerate}
\item Structural Fingerprinting
\item Candidate Cohort Construction
\item Semantic-Assisted Cohort Refinement
\item Cohort-Level Semantic Role Inference
\item Primitive-Level Role Grounding
\end{enumerate}

\subsection{Stage 1: Structural Fingerprinting}
The first stage constructs structural fingerprints for every graphical primitive in the SVG. Its goal is to convert the heterogeneous SVG representation into a structured description of rendered geometry that exposes measurable signals about how each element appears in the visualization. These fingerprints form the basis for cohort construction in the next stage and provide structural cues throughout the pipeline.
Direct reasoning over raw SVG markup is unreliable because rendered geometry often depends on factors not explicitly encoded in local attributes. Nested transformations, inherited style properties, and grouping structures can alter an element's appearance without modifying its local markup. As a result, primitives that appear visually similar may differ syntactically in the SVG, while primitives with similar markup may occupy very different positions in the rendered chart.

To resolve this ambiguity, primitives are analyzed in their rendered form rather than through static markup alone. The SVG is evaluated in a DOM-based rendering environment where layout-dependent properties such as bounding boxes and computed styles can be obtained. This makes it possible to recover the effective geometry and visual styling of each element as it appears in the rendered visualization.

Each graphical primitive is then inspected and recorded in an element index containing geometric, stylistic, and structural descriptors. In particular, for each element $e_i$, the following descriptors are derived.

\begin{itemize}
\item the primitive tag type (e.g., \texttt{rect}, \texttt{path}, \texttt{line}, \texttt{text}),
\item bounding box coordinates obtained via \texttt{getBBox()},
\item width and height of the bounding box,
\item centroid coordinates derived from bounding box geometry,
\item aspect ratio and orientation estimates,
\item computed style attributes including fill, stroke, and opacity,
\item parent group identifier and DOM hierarchy position.
\end{itemize}

These descriptors normalize heterogeneous SVG encodings into a unified geometric representation. For example, primitives positioned through nested transformation matrices are converted into consistent coordinate-space measurements by evaluating their rendered bounding boxes. Similarly, inherited paint attributes are resolved through computed style evaluation so that styling differences between primitives become explicit.

Beyond supporting element-level comparison, these fingerprints also expose higher-level structural cues about the visualization as a whole. Patterns in primitive orientation, spatial distribution, and repetition often reveal global chart structure, enabling the likely visualization type (e.g., bar chart, histogram, scatterplot) to be inferred. This global context later helps constrain the semantic interpretations considered for individual cohorts.

Each primitive is therefore represented by a structural fingerprint

\[
F_i = \langle t_i, g_i, s_i, h_i \rangle
\]

where $t_i$ denotes the primitive tag type, $g_i$ captures geometric descriptors, $s_i$ represents styling attributes, and $h_i$ encodes structural context within the SVG hierarchy.

The output of this stage is a structured primitive index

\[
I = \{e_1, e_2, \dots, e_n\}
\]

in which every SVG element is associated with its structural fingerprint. These fingerprints provide the signals needed to identify repeated patterns, construct candidate cohorts, and establish the structural context for later semantic inference.

\subsection{Stage 2: Candidate Cohort Construction}

Having computed structural fingerprints for every primitive, the next stage infers \texttt{candidate cohorts}, sets of primitives that likely correspond to repeated structural elements within the visualization.

Semantic visualization components rarely correspond to single primitives. Structures such as bar series, scatterplot points, gridlines, or tick labels typically appear as repeated graphical patterns composed of multiple primitives. Identifying these patterns reduces the semantic inference problem from hundreds of heterogeneous elements to a smaller number of structurally coherent groups.

The structural fingerprints computed in Stage~1 enable candidate cohort construction through a sequence of structural analyses that detect repeated visual patterns in the rendered geometry. In practice, cohorts are inferred by identifying groups of primitives that exhibit compatible structural fingerprints and participate in consistent spatial arrangements within the visualization.

\paragraph{Cohort Definition.}
Let the primitive index produced in Stage~1 be

\[
I = \{e_1, e_2, \dots, e_n\}
\]

where each primitive $e_i$ is associated with a structural fingerprint

\[
F_i = \langle t_i, g_i, s_i, h_i \rangle.
\]

A cohort $C_k$ is defined as a subset of primitives

\[
C_k \subseteq I
\]

whose fingerprints exhibit compatibility under a similarity function

\[
\text{sim}(F_i, F_j).
\]

Intuitively, primitives are likely to belong to the same cohort when their fingerprints share compatible structural characteristics and participate in a consistent visual pattern within the visualization.

\subsubsection{Fingerprint Compatibility Analysis}

The first analysis considers compatibility between primitives based on the components of their structural fingerprints. Primitives whose fingerprints exhibit strong similarity across these components may be grouped into provisional cohorts.

\textbf{Primitive type compatibility.}  
Elements are considered compatible when they share similar SVG tag types (e.g., rectangles with rectangles, paths with paths). This prevents primitives with fundamentally different graphical functions from being grouped prematurely.

\textbf{Geometric similarity.}  
Bounding box dimensions and aspect ratios may be compared to determine whether primitives share consistent shape characteristics. For example, bars in a bar chart typically exhibit similar widths and orientation.

\textbf{Orientation consistency.}  
Orientation may be estimated from bounding box ratios. A primitive may be treated as vertically oriented when

\[
\frac{h_i}{w_i} > \tau_v
\]

and horizontally oriented when

\[
\frac{w_i}{h_i} > \tau_h.
\]

\textbf{Stylistic similarity.}  
Styling attributes contained in the fingerprint (e.g., fill color, stroke color, opacity) may be compared to detect repeated visual encodings.

\textbf{Structural context.}  
The hierarchy component of the fingerprint captures DOM grouping and structural context. Elements sharing similar parent groups or hierarchical positions are more likely to represent related visualization components.

Together, these signals support the inference of provisional cohorts from primitives with compatible structural fingerprints.

\subsubsection{Spatial Repetition Analysis}

The geometric components of the fingerprints also support analysis of spatial repetition patterns that frequently characterize visualization structures.

When primitives exhibit compatible fingerprints, their centroid coordinates and bounding box geometry may be examined to determine whether they form regular spatial arrangements. For example, bar marks in a bar chart often produce a sequence of centroids that are nearly evenly spaced along one axis. Similar patterns may appear in regularly distributed scatterplot points or in sets of parallel line segments spanning the plot region, as commonly observed in gridlines.

Regularity may be detected by examining centroid distributions and identifying clusters with consistent spacing intervals. When such spatial regularities are present, the corresponding primitives are likely to form part of the same repeated structural pattern.

To improve robustness, simple stabilization heuristics may be applied to eliminate weak or accidental groupings. For example, by discarding extremely small clusters or enforcing basic orientation consistency. These heuristics help ensure that candidate cohorts reflect coherent visual structures rather than incidental geometric similarity.

The result of this stage is a set of candidate cohorts

\[
\mathcal{C} = \{C_1, C_2, \dots, C_m\}
\]

where each cohort contains primitives that exhibit compatible structural fingerprints and spatial repetition patterns. 

These cohorts should be understood as \texttt{structural hypotheses} rather than final semantic assignments. In many cases, they align closely with meaningful visualization elements such as bar series, gridlines, or scatterplot points. However, fingerprint similarity alone cannot always uniquely determine the semantic role. Provisional cohorts may therefore contain incidental elements, or semantically related elements may remain split across multiple groups. Subsequent stages refine, reinterpret, and validate these structural hypotheses using contextual and semantic evidence.

\begin{figure}
    \centering
    \includegraphics[width=\linewidth]{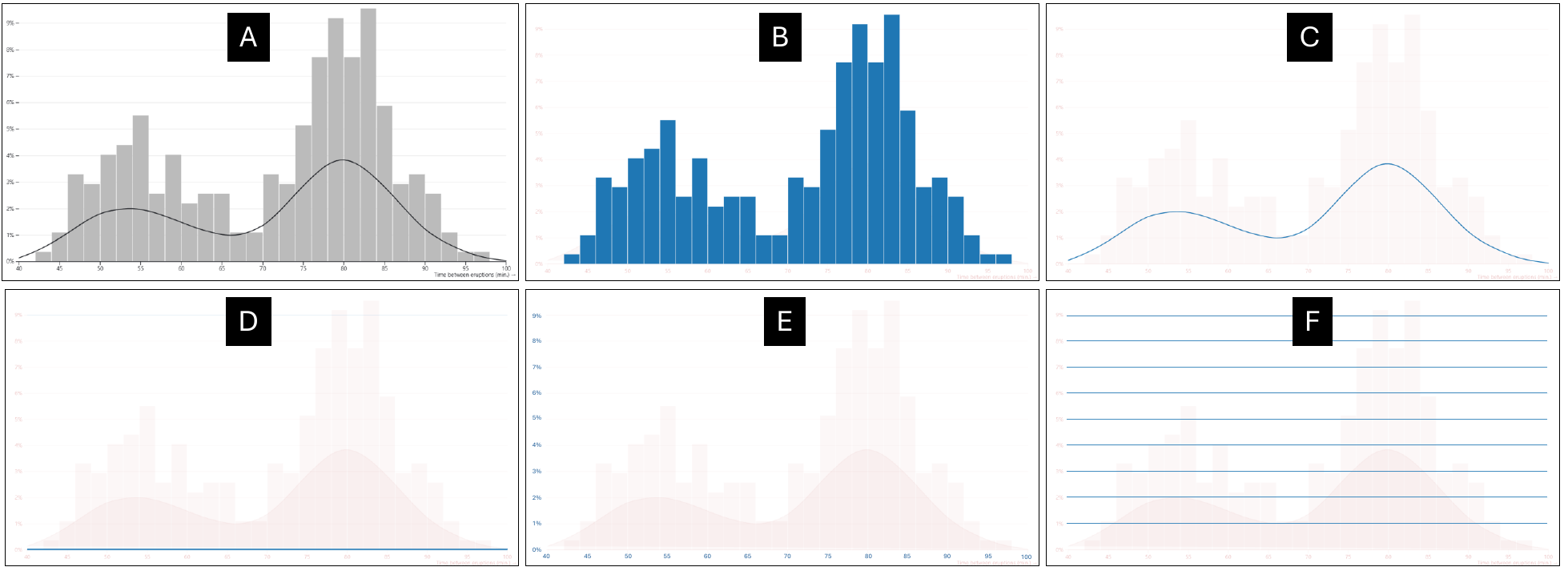}
    \caption{Examples of cohorts generated by an implementation of \CSL. [A] shows the original visualization rendered from SVG (source:D3). [B--F] show constructed cohorts by decomposing [A]. Each cohort encapsulates a subset of SVG elements, representative of a structural subset of visualization: [B$\rightarrow$bars][C$\rightarrow$density curve][D$\rightarrow$x-axis][E$\rightarrow$text][F$\rightarrow$gridlines]. During Stage~4, each cohort is analyzed individually by a multimodal model that infers the semantic roles of the highlighted elements relative to the broader visualization context.  To assist model inference, we render cohort members in blue ink, while the surrounding visualization context is rendered in red ink.}
    \label{fig:cohort-examples}
\end{figure}

\subsection{Stage 3: Semantic-Assisted Cohort Refinement}
Stage~2 infers candidate cohorts from structural fingerprints and spatial regularities observed in the SVG. While these signals capture many repeated patterns, some cohort configurations remain ambiguous. Primitives with similar geometric characteristics may correspond to different semantic roles, and conversely, a single semantic structure may be fragmented across multiple candidate cohorts.

To address such cases, \CSL\  incorporates additional semantic cues that may be present within the visualization. In some SVGs, auxiliary annotations such as ARIA attributes, class names, or identifier patterns provide hints about the chart's intended structure. However, these signals are highly inconsistent across visualization tools and deployment environments and therefore cannot be assumed to be reliably present or consistently meaningful.

Stage~3 therefore treats such signals as evidence for refinement rather than as authoritative labels. A multimodal model analyzes the rendered visualization together with the candidate cohorts and any available SVG annotations, producing contextual judgments about whether these semantic cues convey meaningful structural information.

In effect, this stage assesses the reliability of available semantic channels before incorporating them into the cohort structure. Semantic cues are considered useful when they exhibit consistent patterns across primitives and align with the structural regularities observed in the visualization. Conversely, cues that are sparse, inconsistent, or incompatible with the geometric structure are disregarded. For example, if only a few primitives within an otherwise geometrically consistent cohort contain unrelated class labels or identifier patterns, these annotations are treated as incidental and do not influence refinement.
When semantic cues provide coherent contextual evidence, they may guide two forms of refinement.

One case arises when a candidate cohort exhibits internal heterogeneity, meaning that primitives sharing geometric similarity correspond to distinct semantic roles. For example, short horizontal line segments near an axis may be geometrically similar to fragments of gridlines while serving a different semantic function such as tick marks. In such cases, auxiliary annotations may support partitioning the cohort into smaller semantically coherent subgroups.

Another case arises when multiple candidate cohorts correspond to fragments of the same semantic structure. Minor stylistic differences or grouping conventions may cause repeated marks to be separated during structural analysis. When semantic cues indicate that these fragments represent the same type of chart component, the corresponding cohorts may be merged into a unified group.

Importantly, semantic cues serve only as guidance. Any proposed refinement must remain consistent with the structural evidence derived from the SVG fingerprints. When semantic cues appear ambiguous, inconsistent, or uninformative, they are disregarded and the candidate cohorts produced in Stage~2 remain unchanged.

The result of this stage is a refined cohort set

\[
\mathcal{C}' = \{C'_1, C'_2, \dots, C'_{m'}\},
\]

whose groupings more closely reflect semantically coherent visualization structures while remaining grounded in observable geometry of the SVG. These refined cohorts provide structural units for the subsequent stage, where semantic roles are inferred for each cohort. Figure~\ref{fig:cohort-examples} shows example cohorts generated by an implementation of \CSL.

\subsection{Stage 4: Cohort-Level Semantic Role Inference}

The goal of Stage~4 is to infer the semantic role associated with each cohort as a whole. In other words, this stage determines which visualization component (e.g., data marks, axis ticks, gridlines) the primitives grouped within a given cohort collectively correspond to.

Direct semantic labeling at the level of individual primitives would require reasoning over hundreds or thousands of heterogeneous elements, which is both computationally inefficient and prone to ambiguity. Instead, \CSL\  performs semantic interpretation at the cohort level. Because each cohort groups primitives that exhibit compatible geometric, stylistic, and structural properties, it provides a more stable unit for semantic reasoning.

Semantic roles are inferred through multimodal analysis of the visualization context \cite{wang2024text}. Rather than interpreting the entire visualization at once, \CSL\  presents the model with visual highlights of the primitives belonging to each cohort together with the surrounding chart context (Figure~\ref{fig:cohort-examples}). This decomposition enables the model to reason about each highlighted subset as a structurally coherent candidate component situated within the broader visualization.

Restricting interpretation to cohort-level visual structures reduces task complexity. Instead of analyzing hundreds of heterogeneous primitives simultaneously, the model focuses on smaller sets of visually related elements that likely correspond to a single chart component. In this way, cohort decomposition acts as a structural prior that guides vision-language reasoning toward semantically meaningful visualization units.

Interpreting the semantic role of a cohort often depends on the broader structure of the visualization. The model therefore considers both the local geometric properties of the cohort and global chart context, such as axis layout, mark repetition patterns, and the inferred visualization type, to disambiguate semantic roles.
For example, consider cohort~B in Figure~\ref{fig:cohort-examples}. The highlighted primitives form a set of vertically oriented rectangles with consistent widths and regularly spaced centroids along the horizontal axis. When analyzed in the context of the surrounding visualization structure, including the presence of a quantitative x-axis and regularly spaced rectangular marks, the model may infer that this cohort corresponds to histogram bins. The resulting interpretation assigns the taxonomy-level role \texttt{data\_mark} together with a more specific visualization-part interpretation such as \texttt{histogram\_bin}.

Semantic interpretation therefore produces two complementary outputs. First, a \texttt{taxonomy-level role} identifies the general functional category of the cohort (e.g., \texttt{data\_mark}, \texttt{axis\_tick}, \texttt{gridline}). Second, the model may infer a more specific \texttt{visualization-part interpretation} describing the concrete structural form of the cohort (e.g., bar, scatter point, histogram bin, density curve). This additional level of interpretation provides more precise structural context while remaining compatible with the controlled role taxonomy.

Table~\ref{tbl:mark-taxonomy} presents an initial taxonomy of common visualization roles frequently observed in data visualizations. This taxonomy constrains the semantic search space to interpretable visualization components. It is not intended to be exhaustive and can be extended as additional visualization structures are encountered.

\begin{table}[t]
\centering
\small
\renewcommand{\arraystretch}{1.2}
\begin{tabularx}{\columnwidth}{lX}
\hline
\textbf{Role} & \textbf{Description} \\
\hline
Data\_mark & Graphical marks encoding data values (e.g., bars, points) \\
Axis\_tick & Tick marks indicating scale positions on an axis \\
Tick\_label & Text labels associated with axis ticks \\
Gridline & Reference lines spanning the chart region \\
Legend\_entry & Symbols or text representing legend items \\
Annotation & Textual or graphical explanatory elements \\
\hline
\end{tabularx}
\caption{Initial taxonomy of visualization roles used to guide cohort-level semantic inference. The taxonomy constrains the role space considered by the model during interpretation, allowing assignments to be expressed using a consistent, interpretable vocabulary of common visualization components.}
\label{tbl:mark-taxonomy}
\end{table}

In addition to assigning a role and visualization-part interpretation, cohort-level inference may produce auxiliary semantic hints that support downstream grounding. These hints describe structural expectations associated with the inferred role. For example, the model may indicate expected primitive families (e.g., \texttt{rect\_like} or \texttt{point\_like}) or suggest candidate encoding channels such as vertical position, horizontal position, or color. Such information does not constitute a definitive interpretation but instead provides contextual evidence that can guide the deterministic procedures applied in Stage~5.

Constraining semantic inference to a predefined taxonomy also enables the use of structured output representations that improve robustness. Rather than producing unrestricted free text, semantic reasoning may be guided by schemas that require roles to be selected from the taxonomy and optionally accompanied by supporting contextual cues. 

The result of this stage is a set of cohort-level semantic descriptions

\[
\mathcal{S} =
\{(C'_1, r_1, h_1), (C'_2, r_2, h_2), \dots, (C'_{m'}, r_{m'}, h_{m'})\},
\]

where each refined cohort \(C'_i\) is associated with a semantic role \(r_i\) and auxiliary semantic hints \(h_i\). These cohort-level descriptions provide the input for the final stage of the pipeline, where the inferred roles are deterministically grounded in the SVG primitives belonging to each cohort and incorporated into the resulting Semantic SVG representation.

\subsection{Stage 5: Primitive-Level Role Grounding}

Primitive-level grounding translates cohort-level semantic interpretations inferred in Stage~4 into concrete assignments between inferred semantics and individual SVG primitives. This step relies on deterministic structural criteria derived from the geometric and spatial properties of the SVG elements. These criteria reflect common regularities in visualization design and allow cohort-level interpretations to be resolved at the level of individual primitives.

The auxiliary hints produced in Stage~4 provide additional context for this grounding process. These hints describe structural expectations associated with the inferred role and visualization-part interpretation. For example, match hints describing expected primitive families (e.g., \texttt{rect\_like} or \texttt{point\_like}) may indicate which shapes within a cohort are likely to realize the inferred component, while encoding hints may suggest which geometric properties, such as vertical position, horizontal position, or color, represent encoded data values. In addition, semantic annotations present in the SVG itself (e.g., ARIA attributes, class labels, or identifier patterns) may provide complementary signals when available. Such information does not override structural evidence but can help narrow the set of plausible candidates. Two forms of grounding can commonly arise: 1) Primary element identification and 2) Associated element resolution. 

\textbf{Primary element identification.}  
Certain roles correspond to graphical primitives that directly implement a visualization component. For instance, when a cohort is interpreted as containing \texttt{axis\_tick} elements, geometric criteria such as short line segments aligned with an axis may identify the primitives representing the tick marks themselves. Similarly, cohorts labeled \texttt{data\_mark} with visualization-part interpretations such as \texttt{bar} or \texttt{scatter\_point} may be grounded to rectangular primitives representing bars or point-like primitives representing scatterplot marks.

\textbf{Associated element resolution.}  
Many visualization components consist of multiple related primitives. Axis ticks may appear together with nearby text labels, legend entries may include both graphical symbols and textual descriptions, and axis structures may include domain lines alongside tick marks. Grounding therefore identifies not only the primary primitives corresponding to a role but also associated elements that participate in the same structural component.

To maintain semantic consistency, primitive-level assignments must remain compatible with the geometric and spatial properties derived in earlier stages. In addition, each primitive is associated with at most one semantic role, preventing conflicting interpretations across overlapping cohort structures.

When structural evidence contradicts a cohort-level semantic interpretation, the grounding stage may revise or reject that assignment. In such cases, structural consistency takes precedence, ensuring that semantic interpretations remain faithful to the observable geometry of the visualization.

The result of this stage is a primitive-level semantic mapping

\[
\mathcal{G} =
\{(e_1, m_1, r_1, ...), (e_2, m_2, r_2, ...), \dots, (e_n, m_n, r_n, ...)\},
\]

where each SVG primitive \(e_i\) is associated with a grounded semantic mark \(m_i\),  role \(r_i\), and, where applicable, its corresponding data-role (if element encodes data), axis information, and so on. This mapping forms the basis of the resulting \texttt{Semantic SVG} representation, in which graphical elements are annotated with semantic attributes such as taxonomy-level roles, visualization-part interpretations, cohort membership, and structural relationships.

For example, a rectangular primitive belonging to cohort~B in Figure~\ref{fig:cohort-examples} may appear in the original SVG as:

\begin{lstlisting}
<rect x="142" y="86" width="24" height="110" fill="#4C78A8"/>
\end{lstlisting}

After \CSL\  grounding, the element may be augmented with semantic annotations:

\begin{lstlisting}[
  morekeywords={mark,role,data-role,cohort_id,y_axis,x_axis,match_hint}
]
<rect x="142" y="86" width="24" height="110" fill="#4C78A8"
      mark="area"
      role="histogram_bin"
      data-role="data"
      cohort_id="B"
      y_axis="frequency"
      x_axis="bin_position"
      match_hint="rect_like, vertically oriented"/>
\end{lstlisting}

% \begin{lstlisting}[
%   language=XML,
%   escapeinside={(*}{*)},
%   basicstyle=\ttfamily\small,
%   alsoletter={-,_},
%   morekeywords={mark,role,data-role,cohort_id,y_axis,x_axis,match_hint}
% ]
% <rect x="142" y="86" width="24" height="110" fill="#4C78A8"
%       mark="area"
%       role="histogram_bin"
%       data-role="data" <!-- as opposed to reference/structural -->
%       cohort_id="B"
%       y_axis="frequency"
%       x_axis="bin_position"
%       match_hint="rect_like, vertically oriented"/>
% \end{lstlisting}

% \begin{lstlisting}[
%   language=XML,
%   escapeinside={(*}{*)}, 
%   basicstyle=\ttfamily\small,
%   morekeywords={mark, role, data-role, cohort_id, y_axis, x_axis, match_hint}
% ]
% <rect x="142" y="86" width="24" height="110" fill="#4C78A8"
%       mark="area" 
%       role="histogram_bin"
%       data-role="data" <!--as opposed to reference/structural-->
%       cohort_id="B"
%       y_axis="frequency"
%       x_axis="bin_position"
%       match_hint="rect_like, vertically oriented"/>
% \end{lstlisting}
The resulting Semantic SVG representation exposes visualization elements through explicit semantic roles, structural interpretations, and encoding relationships. This enriched representation forms the basis for subsequent analysis and interaction, and serves as the primary object evaluated in the following section.

\section{Evaluation}
% Our goal is to evaluate \CSL\ as an end-to-end approach for recovering semantic structure from deployed SVG visualizations. To this end, we developed a prototype that implements the full \CSL\  pipeline using the OpenAI GPT-5.4 API and produces SSVG representations from rendered SVG artifacts. Full implementation details, including model prompts, structured response schemas, and the prototype itself, are available in the paper’s OSF repository.

We evaluated \CSL\ as an end-to-end approach for recovering
semantic structure from SVG visualizations. We structured the
evaluation around three critical 
dimensions of post-deployment
semantic recovery. 

\begin{itemize}
    \item \textbf{Feasibility.} Can semantic roles be recovered from deployed
    SVG with useful accuracy across diverse visualizations?

    \item \textbf{Tractability via cohorting.} Does cohort-based decomposition
    materially reduce the difficulty of semantic inference compared to
    whole-chart reasoning?

    \item \textbf{Consistency.} Are the recovered semantic labels consistent
    across repeated runs, indicating stable interpretations rather than
    one-off plausible outputs?
\end{itemize}

To do so, we developed a prototype that implements the full \CSL\  pipeline using the OpenAI GPT-5.4 API and produces SSVG representations from rendered SVG artifacts. 
Full implementation details, including model prompts, structured response schemas, and the prototype itself, are available in the paper’s OSF repository.

\begin{figure}
    \centering
    \includegraphics[width=\linewidth]{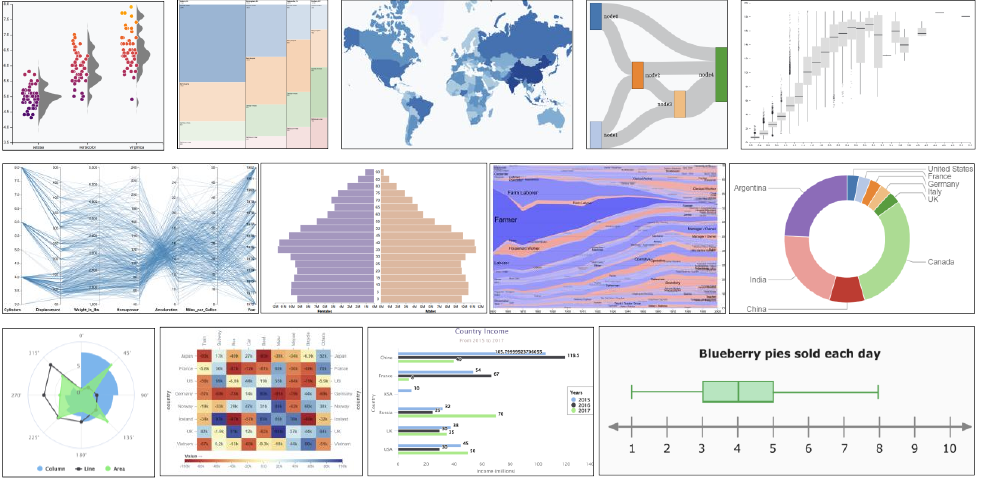}
    \caption{Representative examples from our evaluation corpus of 102 curated SVGs, comprising charts from D3 (top), Vega (middle), and VisAnatomy (bottom)}
    \label{fig:svg-examples}
\end{figure}

\subsection{Evaluation corpus}
We curated a corpus of 102 diverse SVG visualizations drawn from three sources, including D3 Gallery examples (\(n=33\))~\cite{d3graphgallery}, Vega examples (\(n=34\))~\cite{vegaexamples}, and charts from the VisAnatomy dataset (\(n=35\))~\cite{chen2025visanatomy} (Fig.~\ref{fig:svg-examples}). These sources follow different authoring conventions and vary substantially in visual structure, chart composition, and DOM organization. The corpus spans 51 chart types, including statistical distributions, temporal visualizations, part-to-whole displays, geospatial visualizations, hierarchical layouts, and domain-specific forms such as candlestick charts, Gantt charts, and calendar views.

This diversity is important because semantic recovery depends not only on chart type, but also on the heterogeneity of the underlying SVG artifacts. Different visualization ecosystems produce different grouping conventions, primitive reuse patterns, annotation practices, and levels of structural regularity. Evaluating across this diversity allows us to assess recovery under realistic deployment conditions rather than within a single narrowly controlled chart family.
% We curated a corpus of 104 deployed SVG visualizations drawn from three sources: D3 Gallery examples (\(n=32\))~\cite{TODO}, Vega examples (\(n=33\))~\cite{TODO}, and charts from the VisAnatomy dataset (\(n=35\))~\cite{TODO}. These sources use different authoring conventions and exhibit substantial variation in visual structure, chart composition, and DOM organization. The corpus spans 51 chart types, including statistical distributions, temporal visualizations, part-to-whole displays, geospatial visualizations, hierarchical layouts, and domain-specific forms such as candlestick charts, Gantt charts, and calendar views.

% This diversity matters because semantic recovery depends not only on chart type, but also on the heterogeneity of the underlying SVG artifacts. Different authoring ecosystems and visualization forms produce different grouping conventions, primitive reuse patterns, annotation practices, and levels of structural regularity. Evaluating across this diversity allows us to assess recovery under realistic deployment conditions rather than within a single narrowly controlled chart family.

\subsection{Human validation protocol}
\label{subsec:human-validation}
Since no publicly available benchmark currently provides exhaustive element-level ground-truth annotations spanning visualization mark type, functional role, and data-role for deployed SVG visualizations, we evaluated recovered SSVG outputs through human validation rather than direct comparison against a reference corpus. Existing resources, including VisAnatomy~\cite{chen2025visanatomy}, provide valuable semantic annotations, but they do not offer comprehensive element-level labels across all three semantic dimensions required for our evaluation.

Five evaluators, all co-authors with prior data visualization expertise, independently examined each recovered SSVG alongside its rendered visualization and the semantic annotations produced by \CSL, including \texttt{mark}, \texttt{role}, and \texttt{data-role}. Using the rendered visualization as the ground truth, they judged whether each assigned label matched the actual function of the corresponding SVG element in the chart. The evaluation process was designed to minimize subjectivity. Evaluators did not produce subjective judgments through open-ended interpretation, but simply verified whether each assigned label matched the visible chart component. For instance, if an element was labeled as a bar-chart bar, the evaluator checked whether that element indeed corresponded to a bar-chart bar in the rendered visualization. The task is therefore more appropriately characterized as \texttt{constrained verification} than subjective coding, with the visualization itself serving as the ground truth. 

Evaluation proceeded element by element, and each assessed element was assigned to one of three judgment categories:

\begin{itemize}[leftmargin=*, itemsep=1.5ex]

\item \textbf{Correctly labeled:} The element received a semantic label that correctly matched its role. Example:
    \begin{itemize}[label={$\hookrightarrow$}, leftmargin=2em]
        \item \small \texttt{\textcolor{myred}{<line ...>} $\rightarrow$ \textcolor{myred}{mark=}\textcolor{myblue}{"line"}, \textcolor{myred}{role=}\textcolor{myblue}{"Y-axis-tick"}, \textcolor{myred}{data-role=}\textcolor{myblue}{"reference"}}
        \item \footnotesize \textit{Correct labeling for a y-axis tick mark in a graph.}
    \end{itemize}

\item \textbf{Incorrectly labeled:} The assigned label did not match the element's actual role. Example:
    \begin{itemize}[label={$\hookrightarrow$}, leftmargin=2em]
        \item \small \texttt{\textcolor{myred}{<line ...>} $\rightarrow$ \textcolor{myred}{mark=}\textcolor{myblue}{"line"}, \textcolor{myred}{role=}\textcolor{myblue}{"gridline"}, \textcolor{myred}{data-role=}\textcolor{myblue}{"reference"}}
        \item \footnotesize \textit{Incorrect labeling for a y-axis domain interpreted as a gridline.}
    \end{itemize}

\item \textbf{Missed:} A meaningful component was present, but no semantic label was assigned. Example:
    \begin{itemize}[label={$\hookrightarrow$}, leftmargin=2em]
        \item \small \texttt{\textcolor{myred}{<line ...>} $\rightarrow$ \textcolor{myred}{mark=}\textcolor{myblue}{""}, \textcolor{myred}{role=}\textcolor{myblue}{""}, \textcolor{myred}{data-role=}\textcolor{myblue}{""}}
        \item \footnotesize \textit{Missed labeling for a y-axis domain.}
    \end{itemize}

\end{itemize}

To gauge evaluation reliability, a subset of \texttt{15} SSVGs was independently evaluated by two evaluators. Inter-rater reliability on these overlapping cases was high ($\alpha = 0.99$, Krippendorff's alpha), indicating strong agreement in evaluators' assessments.

\subsection{Measurement and Metrics}
\label{subsec:metric}
We derived all reported evaluation metrics from the element-level judgments described in Section~\ref{subsec:human-validation}. These judgments classify each SVG element as correctly labeled, incorrectly labeled, or missed for each semantic attribute (mark, role, and data-role).

% Accuracy computation

To evaluate semantic labeling performance, we used macro-averaged accuracy, computed at the per-SVG level. This choice was motivated by the highly imbalanced distribution of elements within SVG visualizations, where large numbers of repeated data marks can otherwise dominate aggregate metrics. For each SVG and each semantic attribute, we first computed accuracy independently for each semantic class present in that SVG. At the element level, correctly labeled elements were counted as correct, while both incorrectly labeled and missed elements were treated as incorrect. Class-level accuracies were then averaged with equal weight to obtain a per-SVG macro-average. Corpus-level macro-average accuracy was computed by aggregating these per-SVG values across all visualizations. In addition, we reported summary statistics over per-SVG macro-accuracies, including mean, median, and standard deviation, to characterize performance variability across the corpus.

% Tractability comparison

To assess the contribution of cohort-based decomposition, we compared the full CSL pipeline against a baseline variant that performs semantic inference and propagation over the entire SVG without cohorting. Comparisons were conducted at the per-SVG level using macro-averaged accuracy for each semantic attribute. Statistical significance was evaluated using paired tests over per-SVG accuracy values.

% Consistency measurement

To assess labeling consistency, we repeatedly labeled a randomly selected SVG and measured agreement across runs (Table~\ref{tab:lbl:consistency-anslysis}). For each element, we collected all assigned labels across the 100 runs for \texttt{mark}, \texttt{role}, and \texttt{data-role}. Agreement was computed as follows: when all assigned labels were identical, agreement was 100\%. When labels differed, we determined whether variations were semantic or merely lexical (e.g., alternative phrasings of the same concept). Agreement was then defined as the proportion of labels belonging to the largest semantically equivalent group. For example, if an element received the labels ``axis tick label'' (74 runs), ``tick label'' (18 runs), ``x-axis tick label'' (5 runs), and ``gridline'' (3 runs), and the first three were judged semantically equivalent, then the agreement score for that element--attribute pair would be 97\%. This yields a graded measure of agreement that accommodates minor lexical variation in the model's output while preserving substantive disagreement.

\section{Evaluation Results}
\label{sec:results}

\subsection{Labeling feasibility and accuracy assessment}
\label{sec:labeling-acc}

Using the metrics defined in Section~\ref{subsec:metric}, we evaluated labeling performance over 102 SSVGs for three semantic attributes: \texttt{mark}, \texttt{role}, and \texttt{data-role}. The global macro-average accuracies were \(0.822\) for \texttt{mark}, \(0.853\) for \texttt{role}, and \(0.860\) for \texttt{data-role}. At the per-SVG level, the mean macro-accuracies were \(0.857\) for \texttt{mark}, \(0.848\) for \texttt{role}, and \(0.855\) for \texttt{data-role}, with corresponding medians of \(0.905\), \(0.900\), and \(0.900\). These results indicate generally strong labeling performance across the corpus, with \texttt{role} and \texttt{data-role} performing slightly better overall than \texttt{mark}. The higher medians relative to the means further suggest that performance was strong for many SVGs, while a smaller number of more difficult cases lowered the overall average. Overall, these findings indicate that \CSL\ achieved robust semantic labeling accuracy across the corpus.

We also examined omission errors, that is, meaningful visualization components that were missed entirely rather than mislabeled. These errors were not uniformly distributed across element types. They occurred most often in axis-related structures, particularly axis tick labels (241/1831, 13.2\%), axis ticks (147/1619, 9.1\%), gridlines (29/374, 7.8\%), axis domains (20/118, 16.9\%), and axis titles (11/37, 29.7\%). In contrast, omission rates were very low for core data marks such as bars (10/1882, 0.5\%) and points (10/2123, 0.5\%). This suggests that \CSL\ is generally effective at recovering primary data-bearing marks, but is more likely to omit supporting chart scaffolding and annotation-related structures.

\subsection{Tractability assessment}

To isolate the effects of cohorting on the tractability of semantic recovery, we implemented a baseline variant of the \CSL\ prototype (available on OSF) that retains the same overall machinery but removes cohort-bounded processing. In this version, both semantic inference and structural propagation operate over the full SVG rather than within structurally coherent subsets. The model must therefore infer \texttt{mark}, \texttt{role}, and \texttt{data-role} semantics for the entire visualization in a single pass, and the resulting labels are then propagated globally using the same deterministic structural cues based on ancestry and layout. This makes the labeling problem substantially more difficult: semantic ambiguity increases, the space of plausible element-to-role assignments becomes much larger, and errors can propagate across unrelated parts of the chart.

As shown in Table~\ref{tbl:macro-accuracy}, the baseline variant achieves substantially lower performance than the full cohort-based pipeline across all three labeling dimensions. Paired $t$-tests over per-SVG macro-accuracy values confirmed significant differences for all three attributes: mark ($t=21.95$, $p<0.001$, $d=2.17$), role ($t=23.50$, $p<0.001$, $d=2.33$), and data-role ($t=20.85$, $p<0.001$, $d=2.06$). Wilcoxon signed-rank tests further supported these results ($p<0.001$). These findings indicate that cohort-based decomposition is not merely an efficiency-oriented design choice, but a methodological requirement for making semantic recovery tractable. We treat this whole-chart comparison as a coarse ablation of the cohorting mechanism, with more controlled component-wise ablations remaining for future work.

Finally, to examine whether broad SVG-level characteristics help explain labeling accuracy, we analyzed several coarse properties of each artifact, including visualization type, total element count, tag composition, and the proportion of \texttt{<path>} elements. Across these analyses, we did not observe a notable relationship with labeling accuracy. This suggests that performance is less well explained by global SVG characteristics alone and may instead depend on finer-grained structural properties, including how elements are organized during cohort construction. Because these analyses did not yield notable effects, we do not report them in detail here for space reasons; a complete breakdown is available in the paper’s OSF repository.

\begin{table}[t]
\caption{Semantic labeling accuracy across the corpus of 102 SSVGs, covering 17{,}276 labeled elements. ``Macro-average'' reports corpus-level macro-averaged accuracy across semantic classes, whereas ``Average'' and ``SD'' report the mean and standard deviation of per-SVG macro-accuracy values. We compare the full \CSL\ pipeline against a baseline that performs labeling globally over the entire SVG rather than over cohorts. The large performance gap shows that cohorting is central to the method, reducing semantic ambiguity during inference and making label propagation more reliable by constraining it to structurally coherent subsets of the SVG. Boxplots summarize the distribution of per-SVG macro-accuracy values for mark, role, and data-role labeling.}
\label{tbl:macro-accuracy}
\centering
\small
\setlength{\tabcolsep}{4pt}
\renewcommand{\arraystretch}{1.08}
\resizebox{\columnwidth}{!}{%
\begin{tabular}{@{}p{0.19\columnwidth}p{0.23\columnwidth}p{0.18\columnwidth}p{0.18\columnwidth}p{0.14\columnwidth}@{}}
\toprule
\textbf{} & \textbf{Accuracy type} & \textbf{Macro-average} & \textbf{Average} & \textbf{SD} \\
\midrule
\multirow{3}{*}{With cohorts}
& \cellcolor{gray!12}Mark accuracy      & \cellcolor{gray!12}0.8220 & \cellcolor{gray!12}0.8636 & \cellcolor{gray!12}0.1843 \\
& Role accuracy                         & 0.8534 & 0.8541 & 0.1736 \\
& \cellcolor{gray!12}Data-role accuracy & \cellcolor{gray!12}0.8601 & \cellcolor{gray!12}0.8614 & \cellcolor{gray!12}0.1728 \\
\cmidrule(lr){1-5}
\multirow{3}{*}{Without cohorts}
& \cellcolor{gray!12}Mark accuracy      & \cellcolor{gray!12}0.1745 & \cellcolor{gray!12}0.2761 & \cellcolor{gray!12}0.2253 \\
& Role accuracy                         & 0.0951 & 0.1970 & 0.2048 \\
& \cellcolor{gray!12}Data-role accuracy & \cellcolor{gray!12}0.1474 & \cellcolor{gray!12}0.2543 & \cellcolor{gray!12}0.2220 \\
\addlinespace[4pt]
\multicolumn{5}{@{}c@{}}{\includegraphics[width=0.95\columnwidth]{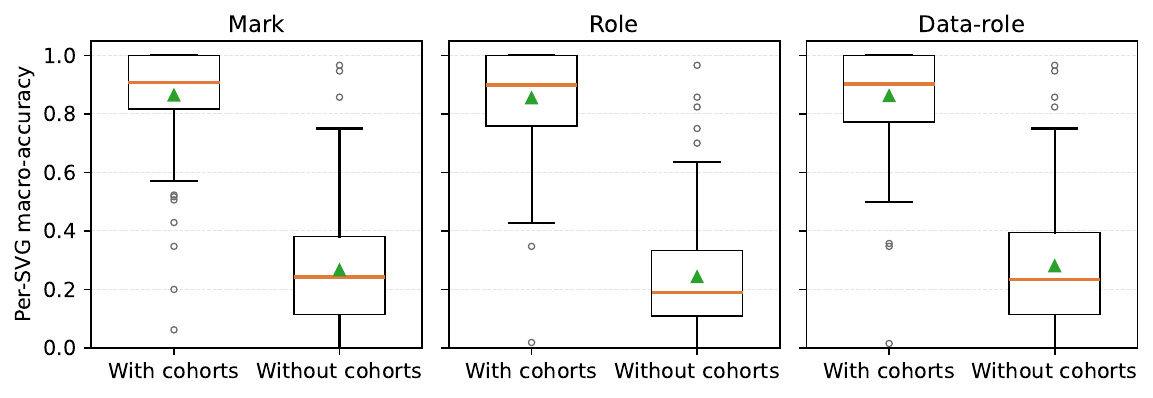}} \\
\bottomrule
\end{tabular}%
}
\end{table}

\subsection{Labeling consistency assessment}
Using the consistency procedure defined in Section~\ref{subsec:metric}, we assessed run-to-run agreement. Across 122 elements of the selected SVG, the mean agreement was \(93.0\%\) for \texttt{mark}, \(91.9\%\) for \texttt{role}, and \(92.7\%\) for \texttt{data-role}, with a median agreement of \(92.0\%\) for all three attributes (Table~\ref{tab:lbl:consistency-anslysis}). Full verbatim agreement was observed for 34 of the 122 shared element IDs (\(27.9\%\)), meaning that all repeated labelings assigned exactly the same value to that element. The remaining 88 IDs showed some degree of variation in wording; however, agreement was assessed as a graded measure based on the proportion of labels falling within the largest semantically equivalent group. These results indicate strong run-to-run stability in the semantic labels produced by \CSL, despite modest lexical variation in some repeated outputs. (Complete details of all analyses reported are available in the paper’s OSF repository.)

\begin{table}[t]
\caption{Labeling consistency was assessed by processing a randomly selected SVG (below) from the study corpus 100 times. The selected graphic consists of 122 drawable elements.}
\label{tab:lbl:consistency-anslysis}
\centering
\small
\setlength{\tabcolsep}{2pt}
\renewcommand{\arraystretch}{1.08}

\begin{tabular*}{\columnwidth}{@{\extracolsep{\fill}}p{0.20\columnwidth}>{\centering\arraybackslash}p{0.70\columnwidth}@{}}
\toprule
\textbf{Metric} & \textbf{Labeling accuracy over 100 runs (mean-median)} \\
\midrule
Mark      & 93.0--92.0\% \\
Role      & 91.9--92.0\% \\
Data-role & 92.7--92.0\% \\
\midrule
\multicolumn{2}{c}{
    \begin{minipage}{\dimexpr\columnwidth-2\tabcolsep}
        \centering
        \includegraphics[width=0.95\linewidth]{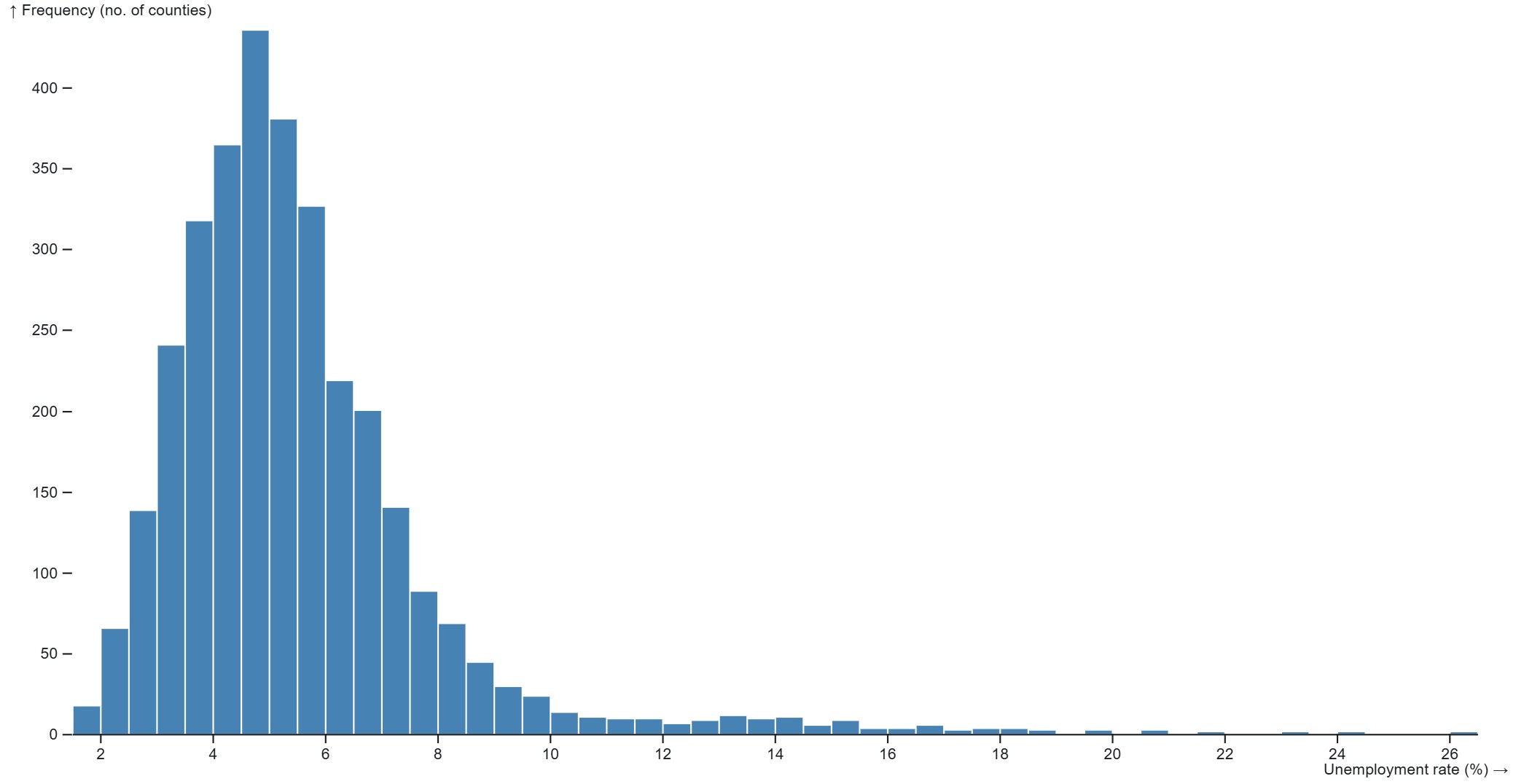}
    \end{minipage}
} \\
\bottomrule
\end{tabular*}
\end{table}
\section{Discussion}

Our results show that \CSL\  can recover semantic structure from deployed SVG visualizations with high accuracy, tractability and consistency across a diverse corpus. In this section, we interpret these findings in terms of the feasibility of post-deployment semantic recovery, the methodological role of cohort-based decomposition, and the downstream implications of recovered semantics for operations such as accessibility augmentation, querying, explanation, personalization, and transformation.

\subsection{Recovering Visualization Semantics as Constrained Inference over Rendered Structure}

Our results suggest that visualization semantics can remain recoverable from the rendered structure itself. Although deployed SVG no longer preserves explicit authoring-time bindings, the strong labeling accuracy across diverse charts and the stability observed across repeated runs indicate that geometric regularities, repetition patterns, and layout conventions retain enough implicit information to support semantic reconstruction \cite{snyder2023divi}. In other words, visualization semantics are not wholly lost after deployment, but remain partially embedded in the spatial and structural organization of the rendered artifact.

A further pattern in our results is that functional semantics appear to be more robustly recoverable than the precise geometric form (Section ~\ref{sec:labeling-acc}).
Across the corpus, role and data-role labeling achieved slightly higher accuracy than mark type (Section~\ref{sec:labeling-acc}), which may suggest that what an element represents in the visualization may be easier to infer than its exact graphical classification. 
This likely reflects the fact that roles are determined not only by local geometry but also by relational context, such as alignment, repetition, and proximity to axes. 
In contrast, mark type distinctions often rely on finer-grained geometric cues that may be more sensitive to variation in SVG encoding (Section \ref{sec:labeling-acc}). This asymmetry suggests that recovering functional structure may be both more stable and especially important for downstream scenarios that depend on identifying what chart elements do, rather than only what they look like \cite{zong2022rich, thompson2023chart, zhang2024charta11y}.

\subsection{Cohort-based decomposition and hybrid-recovery mechanism as foundations for semantic labeling}

A central implication of our findings is that cohort-based decomposition is fundamental to semantic recovery. 
Reasoning directly over hundreds or thousands of heterogeneous SVG primitives is both computationally expensive, unstable, and semantically underconstrained \cite{wu2021ai4vis, chen2023corpora}. By grouping elements into structurally coherent cohorts, \CSL\  reduces the search space, limits interference across unrelated parts of the chart, and provides the model with more interpretable units of inference. In this sense, cohorting does not merely partition the SVG, but introduces a structural scaffold that aligns semantic reasoning with the compositional organization of the visualization \cite{satyanarayan2016vega, liu2024manipulable}.
The ablation study (Table~\ref{tbl:macro-accuracy}) reinforces this conclusion. The performance drop across all three attributes was statistically significant and practically large (Section~\ref{sec:labeling-acc}), confirming that cohorting is not merely an efficiency-oriented design choice but a methodological requirement for making semantic recovery tractable.

Another foundation of semantic recovery is the hybrid mechanism of \CSL. Deployed SVG retains geometric, stylistic, and hierarchical signals, but does not preserve explicit semantic bindings \cite{chen2025svgenius, snyder2025challenges}. Recovering semantics from such artifacts therefore requires more than deterministic rules or unconstrained vision-language reasoning alone \cite{harper2014deconstructing, xie2025datawink, chen2025svgenius}. \CSL\  addresses this challenge through a hybrid design that combines model-based inference with deterministic grounding. 
The AI model contributes to interpretive flexibility by proposing candidate semantic labels for structurally coherent cohorts. Subsequently, deterministic procedures provide a structural discipline that validates, refines, and propagates those labels using observable constraints in the rendered artifact. 
Semantic labeling thus becomes both context-sensitive and structurally anchored, making this hybrid mechanism a second key foundation of recovery alongside cohort-based decomposition.

The weak association between coarse SVG-level properties, such as element count, tag composition, and visualization type (Section \ref{sec:labeling-acc}), and labeling accuracy suggests that global SVG characteristics are unlikely to be the strong predictors of recovery success. Instead, labeling accuracy is more likely to depend on the quality of the cohorts that can be constructed from the rendered artifact. What matters most is whether the SVG contains enough grouping cues to support decomposition into semantically coherent cohorts and, in turn, reliable cohort-based inference and grounding.

\subsection{Implications for downstream operations}
A broader implication of our findings is that deployed SVG can be transformed from a terminal graphical artifact into SSVG, a semantically enriched representation.
If semantic structure can be recovered with high accuracy, tractability and consistency, published visualizations can become substrates for downstream systems that operate over meaningful components rather than raw graphical primitives. This is particularly relevant for accessibility, querying, explanation, personalization, and transformation, where current support often depends on authoring-time semantic structure that might be unavailable after deployment \cite{wimer2025bridging, shin2025drillboards, fan2023accessibility, sharif2021understanding, srinivasan2023azimuth}.

For accessibility, the recovered SSVG structure could help bridge the gap between visually interpretable charts and charts that expose the navigable structure to assistive technologies \cite{elavsky2023data, zong2025semantic}. 
If deployed visualizations can be enriched with recovered semantic roles and structural relationships, downstream tools can target chart components directly rather than relying on brittle manipulations of DOM fragments \cite{seo2024maidr}.

Our findings also suggest an opportunity for future visualization authoring systems to support post-deployment semantic recovery \cite{harper2014deconstructing, snyder2023divi}. 
Although \CSL\  operates without access to authoring-time specifications, recovery may be made easier and more reliable if visualization tools preserve lightweight structural or semantic cues in the deployed SVG. 
Such cues could include a more informative grouping structure, stable identifiers, or optional semantic annotations that do not affect rendering but help expose relationships among chart components \cite{harper2014deconstructing, snyder2023divi}. 

More broadly, \CSL\  reconstructs part of the semantic layer that is typically lost between authoring-time specifications and rendered artifacts. In this sense, SSVG is not merely an annotation format, but a recovered intermediate representation that re-enables forms of interaction, querying, and transformation that would otherwise require access to the original specification \cite{satyanarayan2016vega, satyanarayan2015reactive, wilkinson2011grammar, srinivasan2023azimuth}.

\section{Limitation and future work}

Several limitations should be acknowledged. First, although our corpus spans multiple sources and chart types from D3 \cite{d3graphgallery}, Vega \cite{vegaexamples}, and VisAnatomy \cite{chen2025visanatomy},
it is not exhaustive of the broader space of deployed SVG visualizations. Recovery quality will likely vary across authoring ecosystems, specialized domains, and highly customized hand-written SVGs.
Second, we evaluated \CSL\  using a single commercial multimodal model, OpenAI’s GPT-5.4, for the AI-assisted semantic inference. Our results, therefore, establish the feasibility of the recovery pipeline under a specific model configuration, 
but do not determine how performance varies across different models or scales. This decision was motivated by the primary goal of assessing the fundamental feasibility of \CSL\  rather than benchmarking performance across the broader multimodal model ecosystem. Furthermore, this scope was necessary to focus the research on the internal and external validity of our core assessments.

The present results were obtained without model fine-tuning or systematic use of task-specific in-context learning strategies tailored to semantic labeling. 
Nonetheless, the fact that \CSL\   achieved strong labeling accuracy suggests that the recovery problem is already tractable under a relatively direct prompting setup. This, in turn, makes it plausible that further gains could be achieved through task-specific adaptation, including prompt optimization, retrieval of representative exemplars, or fine-tuning on labeled SVG corpora. Such adaptation may also enable smaller, more efficient models to achieve competitive performance, thereby reducing the practical cost of semantic labeling in terms of speed, compute demand, and associated environmental overhead. Finally, our evaluation scope focused on semantic labeling quality and consistency rather than directly measuring the utility of recovered SSVG in downstream end-user tasks.

Several directions follow from these limitations. Improving cohort construction remains an important next step. Better similarity functions, more adaptive refinement procedures, or learned methods for identifying cohort boundaries could improve recovery in structurally ambiguous visualizations. Supporting structures such as axes, titles, and gridlines also merit more targeted treatment, since they appear more vulnerable to omission and confusion than core data marks. 
Another important direction is systematic evaluation across multiple model families and capability levels to better separate pipeline-level contributions from model-specific effects. Future work should examine task-specific adaptation to improve recovery quality with smaller, more efficient models, while also extending quantitative encoding recovery and semantic coverage, and evaluating the downstream utility of SSVG for applications such as accessibility augmentation, semantic querying, explanation, and user-directed transformation.

\section{Conclusion}

This paper shows that post-deployment semantic recovery becomes tractable through the two complementary foundations introduced by \CSL: (1) cohort-based decomposition and (2) hybrid semantic grounding.
Cohorting reduces the semantic assignment space by organizing heterogeneous SVG primitives into structurally coherent subsets, while the hybrid combination of model-based inference and deterministic grounding makes semantic labeling both context-sensitive and structurally anchored. These mechanisms allow \CSL\  to reconstruct semantic roles, structural relationships, and aspects of quantitative encoding from deployed SVG visualizations with  high accuracy, tractability and consistency across a diverse corpus.
Our findings establish post-deployment semantic recovery as a viable reconstruction problem and position Semantic SVG as a promising intermediate representation for future systems that support accessibility, querying, explanation, personalization, and transformation over published visualizations without requiring access to the original specification.

\bibliographystyle{abbrv-doi-hyperref}
\balance
\bibliography{template}

@misc{d3graphgallery,
  author = {Holtz, Yan},
  title = {D3 Graph Gallery},
  year = {2024},
  howpublished = {\url{https://d3-graph-gallery.com}},
  note = {Accessed: 2026-03-29}
}

@article{bostock2011d3,
  title={D$^3$ data-driven documents},
  author={Bostock, Michael and Ogievetsky, Vadim and Heer, Jeffrey},
  journal={IEEE transactions on visualization and computer graphics},
  volume={17},
  number={12},
  pages={2301--2309},
  year={2011},
  publisher={IEEE}
}

@misc{d3observable,
  author = {Observable},
  title = {D3 Observable Notebooks},
  year = {2026},
  howpublished = {\url{https://observablehq.com/@d3}},
  note = {Accessed: 2026-03-29}
}

@misc{vegaexamples,
  author = {UW Interactive Data Lab},
  title = {Vega Visualization Examples},
  year = {2026},
  howpublished = {\url{https://vega.github.io/vega/examples/}},
  note = {Accessed: 2026-03-29}
}

@article{chen2025visanatomy,
  title={Visanatomy: An svg chart corpus with fine-grained semantic labels},
  author={Chen, Chen and Bako, Hannah K and Yu, Peihong and Hooker, John and Joyal, Jeffrey and Wang, Simon C and Kim, Samuel and Wu, Jessica and Ding, Aoxue and Sandeep, Lara and others},
  journal={IEEE Transactions on Visualization and Computer Graphics},
  year={2025},
  publisher={IEEE}
}

@inproceedings{ying2024vaid,
  title={VAID: Indexing view designs in visual analytics system},
  author={Ying, Lu and Wu, Aoyu and Li, Haotian and Deng, Zikun and Lan, Ji and Wu, Jiang and Wang, Yong and Qu, Huamin and Deng, Dazhen and Wu, Yingcai},
  booktitle={Proceedings of the 2024 CHI Conference on Human Factors in Computing Systems},
  pages={1--15},
  year={2024}
}

@incollection{wilkinson2011grammar,
  title={The grammar of graphics},
  author={Wilkinson, Leland},
  booktitle={Handbook of computational statistics: Concepts and methods},
  pages={375--414},
  year={2011},
  publisher={Springer}
}

@misc{wilkinson2011ggplot2,
  title={ggplot2: elegant graphics for data analysis by WICKHAM, H.},
  author={Wilkinson, Leland},
  year={2011},
  publisher={Oxford University Press}
}

@article{satyanarayan2016vega,
  title={Vega-lite: A grammar of interactive graphics},
  author={Satyanarayan, Arvind and Moritz, Dominik and Wongsuphasawat, Kanit and Heer, Jeffrey},
  journal={IEEE transactions on visualization and computer graphics},
  volume={23},
  number={1},
  pages={341--350},
  year={2016},
  publisher={IEEE}
}

@article{satyanarayan2015reactive,
  title={Reactive vega: A streaming dataflow architecture for declarative interactive visualization},
  author={Satyanarayan, Arvind and Russell, Ryan and Hoffswell, Jane and Heer, Jeffrey},
  journal={IEEE transactions on visualization and computer graphics},
  volume={22},
  number={1},
  pages={659--668},
  year={2015},
  publisher={IEEE}
}

@article{moritz2018formalizing,
  title={Formalizing visualization design knowledge as constraints: Actionable and extensible models in draco},
  author={Moritz, Dominik and Wang, Chenglong and Nelson, Greg L and Lin, Halden and Smith, Adam M and Howe, Bill and Heer, Jeffrey},
  journal={IEEE transactions on visualization and computer graphics},
  volume={25},
  number={1},
  pages={438--448},
  year={2018},
  publisher={IEEE}
}

@article{chen2021vizlinter,
  title={Vizlinter: A linter and fixer framework for data visualization},
  author={Chen, Qing and Sun, Fuling and Xu, Xinyue and Chen, Zui and Wang, Jiazhe and Cao, Nan},
  journal={IEEE transactions on visualization and computer graphics},
  volume={28},
  number={1},
  pages={206--216},
  year={2021},
  publisher={IEEE}
}

@inproceedings{savva2011revision,
  title={Revision: Automated classification, analysis and redesign of chart images},
  author={Savva, Manolis and Kong, Nicholas and Chhajta, Arti and Fei-Fei, Li and Agrawala, Maneesh and Heer, Jeffrey},
  booktitle={Proceedings of the 24th annual ACM symposium on User interface software and technology},
  pages={393--402},
  year={2011}
}

@inproceedings{poco2017reverse,
  title={Reverse-engineering visualizations: Recovering visual encodings from chart images},
  author={Poco, Jorge and Heer, Jeffrey},
  booktitle={Computer graphics forum},
  volume={36},
  number={3},
  pages={353--363},
  year={2017},
  organization={Wiley Online Library}
}

@inproceedings{masson2023chartdetective,
  title={Chartdetective: Easy and accurate interactive data extraction from complex vector charts},
  author={Masson, Damien and Malacria, Sylvain and Vogel, Daniel and Lank, Edward and Casiez, G{\'e}ry},
  booktitle={Proceedings of the 2023 CHI Conference on Human Factors in Computing Systems},
  pages={1--17},
  year={2023}
}

@article{chen2023mystique,
  title={Mystique: Deconstructing svg charts for layout reuse},
  author={Chen, Chen and Lee, Bongshin and Wang, Yunhai and Chang, Yunjeong and Liu, Zhicheng},
  journal={IEEE Transactions on Visualization and Computer Graphics},
  volume={30},
  number={1},
  pages={447--457},
  year={2023},
  publisher={IEEE}
}

@article{ying2024reviving,
  title={Reviving static charts into live charts},
  author={Ying, Lu and Wang, Yun and Li, Haotian and Dou, Shuguang and Zhang, Haidong and Jiang, Xinyang and Qu, Huamin and Wu, Yingcai},
  journal={IEEE Transactions on Visualization and Computer Graphics},
  year={2024},
  publisher={IEEE}
}

@article{liu2024manipulable,
  title={Manipulable semantic components: a computational representation of data visualization scenes},
  author={Liu, Zhicheng and Chen, Chen and Hooker, John},
  journal={IEEE Transactions on Visualization and Computer Graphics},
  volume={31},
  number={1},
  pages={732--742},
  year={2024},
  publisher={IEEE}
}

@techreport{w3c2018waiaria,
  title        = {WAI-ARIA Graphics Module 1.0},
  author       = {Bellamy-Royds, Amelia and Diggs, Joanmarie and Cooper, Michael},
  institution  = {World Wide Web Consortium (W3C)},
  year         = {2018},
  type         = {W3C Recommendation},
  number       = {02 October 2018},
  url          = {https://www.w3.org/TR/graphics-aria-1.0/},
}

@techreport{w3c2018graphicsaam,
  title        = {Graphics Accessibility API Mappings 1.0},
  author       = {Cooper, Michael and Diggs, Joanmarie and Bellamy-Royds, Amelia},
  institution  = {World Wide Web Consortium (W3C)},
  year         = {2018},
  type         = {W3C Recommendation},
  number       = {02 October 2018},
  url          = {https://www.w3.org/TR/graphics-aam-1.0/},
}

@techreport{w3c2026svgaam,
  title        = {SVG Accessibility API Mappings},
  author       = {Shelly, Cynthia and Rogers, Mark},
  institution  = {World Wide Web Consortium (W3C)},
  year         = {2026},
  type         = {W3C Working Draft},
  number       = {20 March 2026},
  url          = {https://www.w3.org/TR/svg-aam-1.0/},
}

@inproceedings{chen2023corpora,
  title={The state of the art in creating visualization corpora for automated chart analysis},
  author={Chen, Chen and Liu, Zhicheng},
  booktitle={Computer Graphics Forum},
  volume={42},
  number={3},
  pages={449--470},
  year={2023},
  organization={Wiley Online Library}
}

@inproceedings{duarte2024autovizua11y,
  title={AutoVizuA11y: A tool to automate screen reader accessibility in charts},
  author={Duarte, Diogo and Costa, Rita and Bizarro, Pedro and Duarte, Carlos},
  booktitle={Computer Graphics Forum},
  volume={43},
  number={3},
  pages={e15099},
  year={2024},
  organization={Wiley Online Library}
}

@inproceedings{andrews2024achart,
  title={Accessible SVG Charts with AChart},
  author={Andrews, Keith and Kopel, Christopher Alexander},
  booktitle={2024 1st Workshop on Accessible Data Visualization (AccessViz)},
  pages={5--8},
  year={2024},
  organization={IEEE}
}

@article{hoque2020d3search,
  title={Searching the visual style and structure of d3 visualizations},
  author={Hoque, Enamul and Agrawala, Maneesh},
  journal={IEEE transactions on visualization and computer graphics},
  volume={26},
  number={1},
  pages={1236--1245},
  year={2019},
  publisher={IEEE}
}

@inproceedings{chen2025tactilevegalite,
  title={Tactile vega-lite: Rapidly prototyping tactile charts with smart defaults},
  author={Chen, Mengzhu and Pedraza Pineros, Isabella and Satyanarayan, Arvind and Zong, Jonathan},
  booktitle={Proceedings of the 2025 CHI Conference on Human Factors in Computing Systems},
  pages={1--23},
  year={2025}
}

@inproceedings{moured2024chartformer,
  title={Chartformer: A large vision language model for converting chart images into tactile accessible svgs},
  author={Moured, Omar and Alzalabny, Sara and Osman, Anas and Schwarz, Thorsten and M{\"u}ller, Karin and Stiefelhagen, Rainer},
  booktitle={International Conference on Computers Helping People with Special Needs},
  pages={299--305},
  year={2024},
  organization={Springer}
}

@article{xie2025datawink,
  title={DataWink: Reusing and Adapting SVG-based Visualization Examples with Large Multimodal Models},
  author={Xie, Liwenhan and Lin, Yanna and Liu, Can and Qu, Huamin and Shu, Xinhuan},
  journal={IEEE Transactions on Visualization and Computer Graphics},
  year={2025},
  publisher={IEEE}
}

@article{snyder2025challenges,
  title={Challenges \& Opportunities with LLM-Assisted Visualization Retargeting},
  author={Snyder, Luke S and Wang, Chenglong and Drucker, Steven M},
  journal={2025 IEEE Visualization and Visual Analytics (VIS)},
  pages={141--145},
  year={2025},
  publisher={IEEE}
}

@inproceedings{lin2025observable,
  title={How Do Observable Users Decompose D3 Code? A Qualitative Study},
  author={Lin, Melissa and Patel, Heer and Lamkin, Medina and Bako, Hannah and Battle, Leilani},
  booktitle={2025 IEEE Visualization and Visual Analytics (VIS)},
  pages={221--225},
  year={2025},
  organization={IEEE}
}

@inproceedings{zong2022rich,
  title={Rich screen reader experiences for accessible data visualization},
  author={Zong, Jonathan and Lee, Crystal and Lundgard, Alan and Jang, JiWoong and Hajas, Daniel and Satyanarayan, Arvind},
  booktitle={Computer Graphics Forum},
  volume={41},
  number={3},
  pages={15--27},
  year={2022},
  organization={Wiley Online Library}
}

@inproceedings{thompson2023chart,
  title={Chart reader: Accessible visualization experiences designed with screen reader users},
  author={Thompson, John R and Martinez, Jesse J and Sarikaya, Alper and Cutrell, Edward and Lee, Bongshin},
  booktitle={Proceedings of the 2023 CHI Conference on Human Factors in Computing Systems},
  pages={1--18},
  year={2023}
}

@inproceedings{jones2024customization,
  title={“Customization is Key”: Reconfigurable Textual Tokens for Accessible Data Visualizations},
  author={Jones, Shuli and Pedraza Pineros, Isabella and Hajas, Daniel and Zong, Jonathan and Satyanarayan, Arvind},
  booktitle={Proceedings of the 2024 CHI Conference on Human Factors in Computing Systems},
  pages={1--14},
  year={2024}
}

@inproceedings{alam2023seechart,
  title={Seechart: Enabling accessible visualizations through interactive natural language interface for people with visual impairments},
  author={Alam, Md Zubair Ibne and Islam, Shehnaz and Hoque, Enamul},
  booktitle={Proceedings of the 28th International Conference on Intelligent User Interfaces},
  pages={46--64},
  year={2023}
}

@inproceedings{battle2018beagle,
  title={Beagle: Automated extraction and interpretation of visualizations from the web},
  author={Battle, Leilani and Duan, Peitong and Miranda, Zachery and Mukusheva, Dana and Chang, Remco and Stonebraker, Michael},
  booktitle={Proceedings of the 2018 CHI conference on human factors in computing systems},
  pages={1--8},
  year={2018}
}

@article{narechania2020nl4dv,
  title={NL4DV: A toolkit for generating analytic specifications for data visualization from natural language queries},
  author={Narechania, Arpit and Srinivasan, Arjun and Stasko, John},
  journal={IEEE Transactions on Visualization and Computer Graphics},
  volume={27},
  number={2},
  pages={369--379},
  year={2020},
  publisher={IEEE}
}

@inproceedings{masry2022chartqa,
  title={Chartqa: A benchmark for question answering about charts with visual and logical reasoning},
  author={Masry, Ahmed and Do, Xuan Long and Tan, Jia Qing and Joty, Shafiq and Hoque, Enamul},
  booktitle={Findings of the association for computational linguistics: ACL 2022},
  pages={2263--2279},
  year={2022}
}

@inproceedings{methani2020plotqa,
  title={Plotqa: Reasoning over scientific plots},
  author={Methani, Nitesh and Ganguly, Pritha and Khapra, Mitesh M and Kumar, Pratyush},
  booktitle={Proceedings of the ieee/cvf winter conference on applications of computer vision},
  pages={1527--1536},
  year={2020}
}

@inproceedings{liu2023deplot,
  title={DePlot: One-shot visual language reasoning by plot-to-table translation},
  author={Liu, Fangyu and Eisenschlos, Julian and Piccinno, Francesco and Krichene, Syrine and Pang, Chenxi and Lee, Kenton and Joshi, Mandar and Chen, Wenhu and Collier, Nigel and Altun, Yasemin},
  booktitle={Findings of the Association for Computational Linguistics: ACL 2023},
  pages={10381--10399},
  year={2023}
}

@article{wang2025internsvg,
    title={InternSVG: Towards Unified SVG Tasks with Multimodal Large Language Models},
    author={Wang, Haomin and Yin, Jinhui and Wei, Qi and Zeng, Wenguang and Gu, Lixin and Ye, Shenglong and Gao, Zhangwei
    and Wang, Yaohui and Zhang, Yanting and Li, Yuanqi and others},
    journal={arXiv preprint arXiv:2510.11341},
    year={2025}
}

@inproceedings{chen2025svgenius,
  title={Svgenius: Benchmarking llms in svg understanding, editing and generation},
  author={Chen, Siqi and Dong, Xinyu and Xu, Haolei and Wu, Xingyu and Tang, Fei and Zhang, Hang and Yan, Yuchen and Wu, Linjuan and Zhang, Wenqi and Hou, Guiyang and others},
  booktitle={Proceedings of the 33rd ACM International Conference on Multimedia},
  pages={13289--13296},
  year={2025}
}

@inproceedings{vaithilingam2024dynavis,
  title={Dynavis: Dynamically synthesized ui widgets for visualization editing},
  author={Vaithilingam, Priyan and Glassman, Elena L and Inala, Jeevana Priya and Wang, Chenglong},
  booktitle={Proceedings of the 2024 CHI Conference on Human Factors in Computing Systems},
  pages={1--17},
  year={2024}
}

@article{liu2024spatial,
  title={A spatial constraint model for manipulating static visualizations},
  author={Liu, Can and Zhang, Yu and Wu, Cong and Li, Chen and Yuan, Xiaoru},
  journal={ACM Transactions on Interactive Intelligent Systems},
  volume={14},
  number={2},
  pages={1--29},
  year={2024},
  publisher={ACM New York, NY}
}

@article{snyder2023divi,
  title={Divi: Dynamically interactive visualization},
  author={Snyder, Luke S and Heer, Jeffrey},
  journal={IEEE Transactions on Visualization and Computer Graphics},
  volume={30},
  number={1},
  pages={403--413},
  year={2023},
  publisher={IEEE}
}

@inproceedings{harper2014deconstructing,
  title={Deconstructing and restyling D3 visualizations},
  author={Harper, Jonathan and Agrawala, Maneesh},
  booktitle={Proceedings of the 27th annual ACM symposium on User interface software and technology},
  pages={253--262},
  year={2014}
}

@article{fu2020chartem,
  title={Chartem: Reviving chart images with data embedding},
  author={Fu, Jiayun and Zhu, Bin and Cui, Weiwei and Ge, Song and Wang, Yun and Zhang, Haidong and Huang, He and Tang, Yuanyuan and Zhang, Dongmei and Ma, Xiaojing},
  journal={IEEE Transactions on Visualization and Computer Graphics},
  volume={27},
  number={2},
  pages={337--346},
  year={2020},
  publisher={IEEE}
}

@article{wu2021ai4vis,
  title={Ai4vis: Survey on artificial intelligence approaches for data visualization},
  author={Wu, Aoyu and Wang, Yun and Shu, Xinhuan and Moritz, Dominik and Cui, Weiwei and Zhang, Haidong and Zhang, Dongmei and Qu, Huamin},
  journal={IEEE Transactions on Visualization and Computer Graphics},
  volume={28},
  number={12},
  pages={5049--5070},
  year={2021},
  publisher={IEEE}
}

@article{dou2024hierarchically,
  title={Hierarchically recognizing vector graphics and a new chart-based vector graphics dataset},
  author={Dou, Shuguang and Jiang, Xinyang and Liu, Lu and Ying, Lu and Shan, Caihua and Shen, Yifei and Dong, Xuanyi and Wang, Yun and Li, Dongsheng and Zhao, Cairong},
  journal={IEEE Transactions on Pattern Analysis and Machine Intelligence},
  volume={46},
  number={12},
  pages={7556--7573},
  year={2024},
  publisher={IEEE}
}

@article{lee2025svg,
  title={SVG Decomposition for Enhancing Large Multimodal Models Visualization Comprehension: A Study with Floor Plans},
  author={Lee, Jeongah and Sarvghad, Ali},
  journal={arXiv preprint arXiv:2511.03478},
  year={2025}
}

@inproceedings{li2022structure,
  title={Structure-aware visualization retrieval},
  author={Li, Haotian and Wang, Yong and Wu, Aoyu and Wei, Huan and Qu, Huamin},
  booktitle={Proceedings of the 2022 CHI Conference on Human Factors in Computing Systems},
  pages={1--14},
  year={2022}
}

@article{wickham2010layered,
  title={A layered grammar of graphics},
  author={Wickham, Hadley},
  journal={Journal of computational and graphical statistics},
  volume={19},
  number={1},
  pages={3--28},
  year={2010},
  publisher={Taylor \& Francis}
}

@inproceedings{wimer2025bridging,
  title={Bridging Chart Extraction and Accessibility in Data Visualization},
  author={Wimer, Brianna},
  booktitle={2025 IEEE Workshop on Accessible Data Visualization (AccessViz)},
  pages={25--29},
  year={2025},
  organization={IEEE}
}

@article{shin2025drillboards,
  title={Drillboards: Adaptive visualization dashboards for dynamic personalization of visualization experiences},
  author={Shin, Sungbok and Na, Inyoup and Elmqvist, Niklas},
  journal={IEEE Transactions on Visualization and Computer Graphics},
  year={2025},
  publisher={IEEE}
}

@article{elavsky2023data,
  title={Data navigator: an accessibility-centered data navigation toolkit},
  author={Elavsky, Frank and Nadolskis, Lucas and Moritz, Dominik},
  journal={IEEE transactions on visualization and computer graphics},
  volume={30},
  number={1},
  pages={803--813},
  year={2023},
  publisher={IEEE}
}

@article{zong2025semantic,
  title={Semantic Scaffolding: Augmenting Textual Structures with Domain-Specific Groupings for Accessible Data Exploration},
  author={Zong, Jonathan and Pineros, Isabella Pedraza and Chen, Mengzhu Katie and Hajas, Daniel and Satyanarayan, Arvind},
  journal={arXiv preprint arXiv:2506.15883},
  year={2025}
}

@inproceedings{jung2017chartsense,
  title={Chartsense: Interactive data extraction from chart images},
  author={Jung, Daekyoung and Kim, Wonjae and Song, Hyunjoo and Hwang, Jeong-in and Lee, Bongshin and Kim, Bohyoung and Seo, Jinwook},
  booktitle={Proceedings of the 2017 chi conference on human factors in computing systems},
  pages={6706--6717},
  year={2017}
}

@article{fan2023accessibility,
  title={The accessibility of data visualizations on the web for screen reader users: Practices and experiences during covid-19},
  author={Fan, Danyang and Fay Siu, Alexa and Rao, Hrishikesh and Kim, Gene Sung-Ho and Vazquez, Xavier and Greco, Lucy and O'Modhrain, Sile and Follmer, Sean},
  journal={ACM Transactions on Accessible Computing},
  volume={16},
  number={1},
  pages={1--29},
  year={2023},
  publisher={ACM New York, NY}
}

@inproceedings{sharif2021understanding,
  title={Understanding screen-reader users’ experiences with online data visualizations},
  author={Sharif, Ather and Chintalapati, Sanjana Shivani and Wobbrock, Jacob O and Reinecke, Katharina},
  booktitle={Proceedings of the 23rd International ACM SIGACCESS Conference on Computers and Accessibility},
  pages={1--16},
  year={2021}
}

@inproceedings{seo2024maidr,
  title={Maidr meets ai: Exploring multimodal llm-based data visualization interpretation by and with blind and low-vision users},
  author={Seo, JooYoung and Kamath, Sanchita S and Zeidieh, Aziz and Venkatesh, Saairam and McCurry, Sean},
  booktitle={Proceedings of the 26th International ACM SIGACCESS Conference on Computers and Accessibility},
  pages={1--31},
  year={2024}
}

@inproceedings{zhang2024charta11y,
  title={ChartA11y: Designing accessible touch experiences of visualizations with blind smartphone users},
  author={Zhang, Zhuohao and Thompson, John R and Shah, Aditi and Agrawal, Manish and Sarikaya, Alper and Wobbrock, Jacob O and Cutrell, Edward and Lee, Bongshin},
  booktitle={Proceedings of the 26th International ACM SIGACCESS Conference on Computers and Accessibility},
  pages={1--15},
  year={2024}
}

@inproceedings{srinivasan2023azimuth,
  title={Azimuth: Designing accessible dashboards for screen reader users},
  author={Srinivasan, Arjun and Harshbarger, Tim and Hilliker, Darrell and Mankoff, Jennifer},
  booktitle={Proceedings of the 25th International ACM SIGACCESS Conference on Computers and Accessibility},
  pages={1--16},
  year={2023}
}

@article{wang2024text,
  title={Text-based reasoning about vector graphics},
  author={Wang, Zhenhailong and Hsu, Joy and Wang, Xingyao and Huang, Kuan-Hao and Li, Manling and Wu, Jiajun and Ji, Heng},
  journal={arXiv preprint arXiv:2404.06479},
  volume={6},
  year={2024}
}

@article{deng2022visimages,
  title={VisImages: A fine-grained expert-annotated visualization dataset},
  author={Deng, Dazhen and Wu, Yihong and Shu, Xinhuan and Wu, Jiang and Fu, Siwei and Cui, Weiwei and Wu, Yingcai},
  journal={IEEE Transactions on Visualization and Computer Graphics},
  volume={29},
  number={7},
  pages={3298--3311},
  year={2022},
  publisher={IEEE}
}

@article{satyanarayan2019critical,
  title={Critical reflections on visualization authoring systems},
  author={Satyanarayan, Arvind and Lee, Bongshin and Ren, Donghao and Heer, Jeffrey and Stasko, John and Thompson, John and Brehmer, Matthew and Liu, Zhicheng},
  journal={IEEE transactions on visualization and computer graphics},
  volume={26},
  number={1},
  pages={461--471},
  year={2019},
  publisher={IEEE}
}

@article{cui2021mixed,
  title={A mixed-initiative approach to reusing infographic charts},
  author={Cui, Weiwei and Wang, Jinpeng and Huang, He and Wang, Yun and Lin, Chin-Yew and Zhang, Haidong and Zhang, Dongmei},
  journal={IEEE Transactions on Visualization and Computer Graphics},
  volume={28},
  number={1},
  pages={173--183},
  year={2021},
  publisher={IEEE}
}

@article{harper2017converting,
  title={Converting basic D3 charts into reusable style templates},
  author={Harper, Jonathan and Agrawala, Maneesh},
  journal={IEEE transactions on visualization and computer graphics},
  volume={24},
  number={3},
  pages={1274--1286},
  year={2017},
  publisher={IEEE}
}

@article{zhu2019towards,
  title={Towards automated infographic design: Deep learning-based auto-extraction of extensible timeline},
  author={Zhu-Tian, Chen and Wang, Yun and Wang, Qianwen and Wang, Yong and Qu, Huamin},
  journal={IEEE transactions on visualization and computer graphics},
  volume={26},
  number={1},
  pages={917--926},
  year={2019},
  publisher={IEEE}
}

\clearpage
\onecolumn
\appendix % You can use the `hideappendix` class option to skip everything after \appendix
% \crefalias{section}{appendix} % this is to make sure that cleverref switches to referring to Appx. X from here on

% \input{new-sections/7-Appendix}

\end{document}